\documentclass[twocolumn]{aastex6}
\usepackage{times}
\usepackage{amsmath}
\usepackage{textcomp}
\usepackage{amssymb}

\newcommand{\fermi}{{\it Fermi}}

\newcommand{\gm}{$\gamma$}

\slugcomment{submitted to ApJ}

\shorttitle{INOV properties of {\it Fermi} Blazars}
\shortauthors{Paliya et al.}

\begin{document}

\title{Intra-Night Optical Variability Monitoring of {\it Fermi} Blazars: First Results from 1.3 \MakeLowercase{m} J C Bhattacharya Telescope}

\author{Vaidehi S. Paliya$^{1}$, C. S. Stalin$^{2}$, M. Ajello$^{1}$, and A. Kaur$^{1}$} 
\affil{$^1$Department of Physics and Astronomy, Clemson University, Kinard Lab of Physics, Clemson, SC 29634-0978, USA}
\affil{$^2$Indian Institute of Astrophysics, Block II, Koramangala, Bangalore-560034, India}
\email{vpaliya@g.clemson.edu}

\begin{abstract}
We report the first results obtained from our campaign to characterize the intranight-optical variability (INOV) properties of \fermi~detected blazars, using the observations from the recently commissioned 1.3 m J C Bhattacharya telescope (JCBT). During the first run, we were able to observe 17 blazars in the Bessel $R$ filter for $\sim$137 hrs. Using $C$ and scaled $F$-statistics, we quantify the extent of INOV and derive the duty cycle (DC) which is the fraction of time during which a source exhibits a substantial flux variability. We find a high DC of 40\% for BL Lac objects and the flat spectrum radio quasars are relatively less variable (DC $\sim15$\%). However, when estimated for blazars sub-classes, a high DC of $\sim$59\% is found in low synchrotron peaked (LSP) blazars, whereas, intermediate and high synchrotron peaked objects have a low DC of $\sim$11\% and 13\%, respectively. We find evidences about the association of the high amplitude INOV with the $\gamma$-ray flaring state. We also notice a high polarization during the elevated INOV states (for the sources that have polarimetric data available), thus supporting the jet based origin of the observed variability. We plan to enlarge the sample and utilize the time availability from the small telescopes, such as 1.3 m JCBT, to strengthen/verify the results obtained in this work and those existed in the literature.

\end{abstract}


\section{Introduction}{\label{sec:Intro}}
Blazars are a special class of active galactic nuclei (AGN) with their relativistic jets pointed towards the observer \citep[][]{1995PASP..107..803U}. Due to the peculiar orientation of the jet that enhances the radiation because of the relativistic boosting , blazars are known to emit over entire electromagnetic spectrum, from low energy radio waves to very high energy \gm-rays. In addition to that, another defining characteristic feature displayed by blazars is the rapid and high amplitude flux variations observed at all the accessible wavelengths \citep[see, e.g.,][]{2002BASI...30..765S,2004MNRAS.348..176S,2010ApJ...715..362J,2011ApJ...730L...8A,2012MNRAS.424.2625B,2015ApJ...811..143P,2016A&A...591A..21M}. Based on the equivalent width (EW) of the optical emission lines, blazars are classified as flat spectrum radio quasars (FSRQs) and BL Lac objects with FSRQs exhibit broad emission lines (EW$>$5\AA). The observation of strong emission lines from FSRQs indicates for the presence of a luminous broad line region (BLR), which in turn, suggests a high and efficient accretion process that illuminate the BLR \citep[e.g.,][]{2012MNRAS.421.1764S}. In fact, the infrared (IR) to ultraviolet (UV) spectrum of many FSRQs is found to be dominated by thermal emission from the accretion disk \citep[e.g.,][]{2010MNRAS.402..497G}. The presence/absence of narrow emission lines in the optical spectrum of BL Lac objects, on the other hand, hints a relatively low and inefficient accretion and/or the dominance of the non-thermal synchrotron radiation originated from the plasma moving along the relativistic jet. \citet[][]{2010ApJ...716...30A} introduced another blazar classification based on the location of the synchrotron peak in the broadband spectral energy distribution (SED) of the sources. A blazar is known as low synchrotron peaked or LSP if the rest-frame synchrotron peak frequency ($\nu^{\rm peak}_{\rm syn}$) is $\lesssim10^{14}$ Hz. On the other hand, for $10^{14}\lesssim\nu^{\rm peak}_{\rm syn}\lesssim10^{15}$ and $\nu^{\rm peak}_{\rm syn}\gtrsim10^{15}$ Hz, the sources are classified as intermediate synchrotron peaked (ISP) or high synchrotron peaked (HSP) blazars.

The intensity variations in blazars are generally explained with the `shock-in-jet' model \citep[][]{1985ApJ...298..114M} and this suggests that the observed temporal variability is associated with the radiative processes occurring in the AGN jets \citep[e.g.,][]{1995ARA&A..33..163W}. However, the detection of the low amplitude optical variability from radio-quiet \citep[see, e.g.,][and references therein]{2016MNRAS.461..666K} hints that such optical flux variations can also originate from the accretion disk instabilities or perturbations \citep[][]{1993ApJ...406..420M}. In the widely accepted scenario of the blazar emission, the high amplitude, rapid optical variability is associated with the non-thermal jetted synchrotron emission. This is also supported from the observations of the correlated multi-wavelength flux variations \citep[e.g.,][]{2011MNRAS.410..368B,2015ApJ...803...15P} that demands the origin of the optical to $\gamma$-ray emission to be co-spatial. The observations of the high degree of variable optical polarization during the flaring activity of the quasars provide further supporting evidences to it \citep[e.g.,][]{2014PASJ...66..108I,2016ApJ...833...77I}.

The causes responsible for the observed flux variability from blazars are not well understood. In the framework of the one-zone leptonic emission models, a simultaneous multi-wavelength flux enhancement is possible due to injection of the fresh, highly energetic particles into the emission region or/and due to the sudden acceleration of the jet \citep[e.g.,][]{2015ApJ...803...15P}. On the other hand, there is evidence that flaring behavior was seen only at optical/UV energies \citep[e.g.,][]{2013ApJ...763L..11C,2014ApJ...789..143P}. Such anomalous outbursts are explained on the basis of the changes in the magnetic field strength or the location of the emission region \citep[][]{2007MNRAS.382L..82G,2014ApJ...789..143P}.

The intra-night optical variability (INOV), or microvariability (i.e., variations with durations extending from minutes to hours) of blazars has been studied for more than two decades \citep[][]{1989Natur.337..627M,1991AJ....101.1196C,2004MNRAS.348..176S,2012MNRAS.424.2625B,2016A&A...591A..21M}. In fact, the study of the INOV properties using small 1$-$2 m class optical telescopes has been proved quite successful in determining the nature of the short-term optical flux variations in blazars and their connection with the high energy \gm-ray flares \citep[e.g.,][]{2015ApJ...809..130C,2016MNRAS.455..680A}. It is found in recent studies that \gm-ray blazars have faster jets or they have higher relativistic beaming factors compared to their non \gm-ray detected counterparts \citep[e.g.,][]{2010A&A...512A..24S,2015ApJ...810L...9L}. Therefore, it is more likely to detect a relatively high amplitude INOV from \gm-ray emitting blazars, thereby making them an important target to study the fast optical flux variability.

The release of the third catalog of the \gm-ray emitting AGNs detected by the {\it Fermi}-Large Area Telescope \citep[3LAC;][]{2015ApJ...810...14A} represents a uniform, flux-limited sample of \gm-ray emitters, of which blazars forms a major fraction. Motivated by the availability of the \gm-ray information for a large number of blazars and also the availability of telescope time on the recently commissioned 1.3 m J C Bhattacharya Telescope (JCBT) at the Vainu Bappu Observatory (VBO), Kavalur, India, we have started an optical monitoring campaign to systematically study the INOV properties of \gm-ray emitting blazars. In this work, we present the results obtained from the first run of the observing campaign. In Section \ref{sec:sample} we discuss our sample selection procedure and provide a brief introduction of the JCBT in Section \ref{sec:JCBT}. We outline the s.pdf of the adopted data reduction in Section \ref{sec:data} and present the results in Section \ref{sec:results}. The results associated with the individual objects are highlighted in Section \ref{sec:objects} and we briefly discuss and summarize our findings in Section \ref{sec:summary}.

\section{Sample}{\label{sec:sample}}
We start with 3LAC blazars that have a redshift measurements reported in the catalog. We select those sources that are observable from VBO (geographical coordinates: 12$^{\circ}$ 34$^{\prime}$ 0$^{\prime\prime}$ N, 78$^{\circ}$ 50$^{\prime}$ 0$^{\prime\prime}$ E). Since we use a relatively small telescope (1.3 m diameter), we retain only those blazars that have apparent $R$ band magnitude $<18$. We further look into the sky images\footnote{http://www.ledas.ac.uk/DSSimage} and remove those sources where the field is crowded or/and if there are relatively bright stars close to the target object (e.g., 3FGL J1104.4+3812 or Mrk 421). This is required to avoid the CCD saturation for long exposures ($>5$ min). At this stage, we are left with about 100 blazars\footnote{We note that the choice of observing a particular source is subjective and therefore our sample is not complete. It depends on the night sky condition, field of view (uncongested or crowded), and the brightness of nearby stars compared to the target quasar. Therefore, it is likely that we may be able to accommodate those blazars that were excluded from the first observing run.}. Our observing campaign depends on the availability of the telescope time and we adopt an `as and when'  approach to fully utilize the monitoring opportunity. Therefore, the observing duration varies from $\sim$1 hr to $\sim$9 hrs on any particular night. Our first observing run started in 2015 January and ended in 2016 January. During this periods, we were able to observe 17 blazars (7 FSRQs and 10 BL Lac objects) for a total of $\sim$137 hrs in 27 nights. The general properties of these sources are given in Table \ref{tab:basic_info}.

\section{J C Bhattacharya Telescope}
\begin{figure*}
\includegraphics[width=\linewidth]{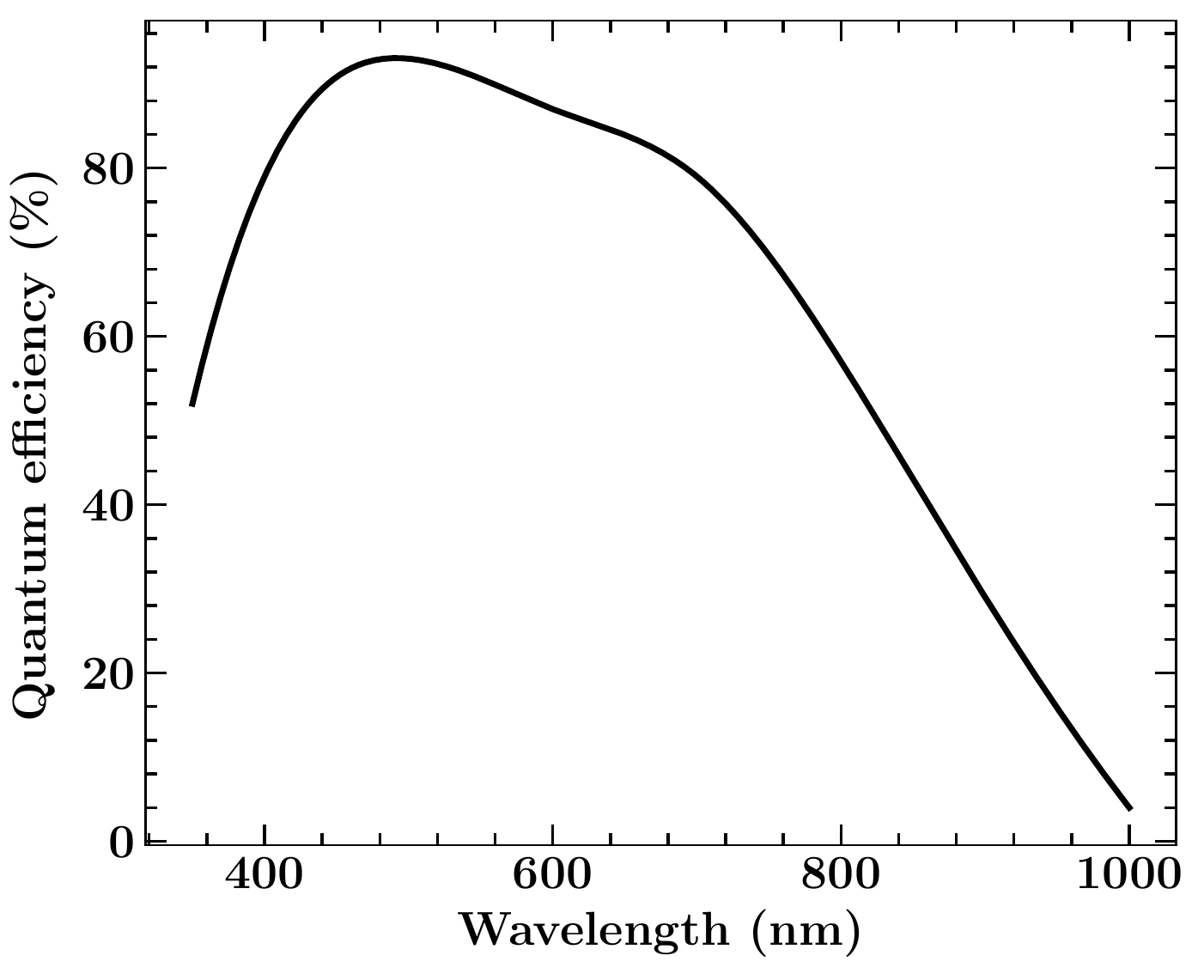}
\caption{The quantum efficiency plot for 2k $\times$ 4k CCD which is used to carry out the observations. We have adopted the Bessel $R$ filter ($\lambda_{\rm eff}\sim 6400$ \AA) for the monitoring campaign.\label{fig_ccd}} 
\end{figure*}

{\label{sec:JCBT}}
The JCBT is a 1.3m F/8 Double Horseshoe telescope at VBO, Tamilnadu, India. It was inaugurated in 2014 April. There are two types of charge coupled devices (CCDs) mounted on the telescope: (i) 1k $\times$ 1k proEM CCD and (ii) 2k $\times$ 4k normal CCD. The proEM CCD is Peltier cooled with water circulation (with Ethylene Glycol mixture) and the liquid Nitrogen is used as a coolent for the normal CCD. The advantage of the ProEM CCD is that the maximum speed is 10 MHz for EM mode and 5 MHz for normal mode, both with variable gain settings. However, it has a small field of view (FOV). Since our primary objective is to perform the differential photometry, we need a suitable pair of non-variable stars present in the same exposure frame. This demands a relatively large FOV and therefore, we choose the 2k $\times$ 4k normal CCD. In all the observations, its central 2k $\times$ 2k region is used. Furthermore, all the observations are carried out using the Bessel $R$ filter ($\lambda_{\rm eff}\sim 6400$ \AA) in which CCD has a high response (see the quantum efficiency plot in Figure \ref{fig_ccd}). The main characteristics of both CCDs are presented in Table \ref{tab:ccd}.
 
\section{Data Reduction}{\label{sec:data}}
We use IRAF\footnote{IRAF is distributed by the National Optical Astronomy Observatories, which is operated by the Association of Universities for Research in Astronomy Inc. under cooperative agreement with the National Science Foundation.} to perform the data reduction. The preliminary cleaning of the image frames is done by preforming the bias subtraction, flat fielding, and cosmic-ray removal. The bias frames are taken at regular intervals throughout the night and flat-field images are taken at dusk, i.e., at the beginning of the observations. We follow the aperture photometry method to extract the instrumental magnitudes of the target blazar and nearby comparison stars present in the same image frame. Among the various comparison stars, we select the pair with the steadiest differential light curve (DLC), i.e., with the minimum standard deviation for a given aperture \citep[taken as full width at half maximum of the point spread function of stars,][]{1989PASP..101..616H}. Once the steadiest pair of stars is determined, the optimum aperture size is derived by considering a range of aperture radii, calculating the standard deviation of the DLC for the selected stars pair and choosing the aperture radius that minimizes the scatter. Using this optimized aperture radius, we extract the instrumental magnitudes of the target blazar and the selected pair of the comparison stars. In Table \ref{tab:com_star}, we report the USNO-B1 cataloged positions and apparent $R$ band magnitudes for the selected comparison stars associated with each blazar. It should be noted that the possible uncertainties in the quoted magnitudes could be up to $\sim$0.25 magnitudes \citep[][]{2003AJ....125..984M}.

The IR-optical spectrum of the bright blazar 3FGL J1653.9+3945 (or Mkn 501) is known to be contaminated from the host galaxy radiation. However, since we are interested in measuring the differential flux rather than the absolute values, we neglect it. Furthermore, the INOV is not expected to be affected from the host galaxy compared to the fast variation originates from the relativistic jet and therefore it is a reasonable choice \citep[see., e.g.,][for a discussion on the host galaxy effect on the long term variability behavior of Mkn 501]{2016ApJS..222...24X}.

\section{Results}
\begin{figure*}
\hbox{
\includegraphics[width=5.5cm]{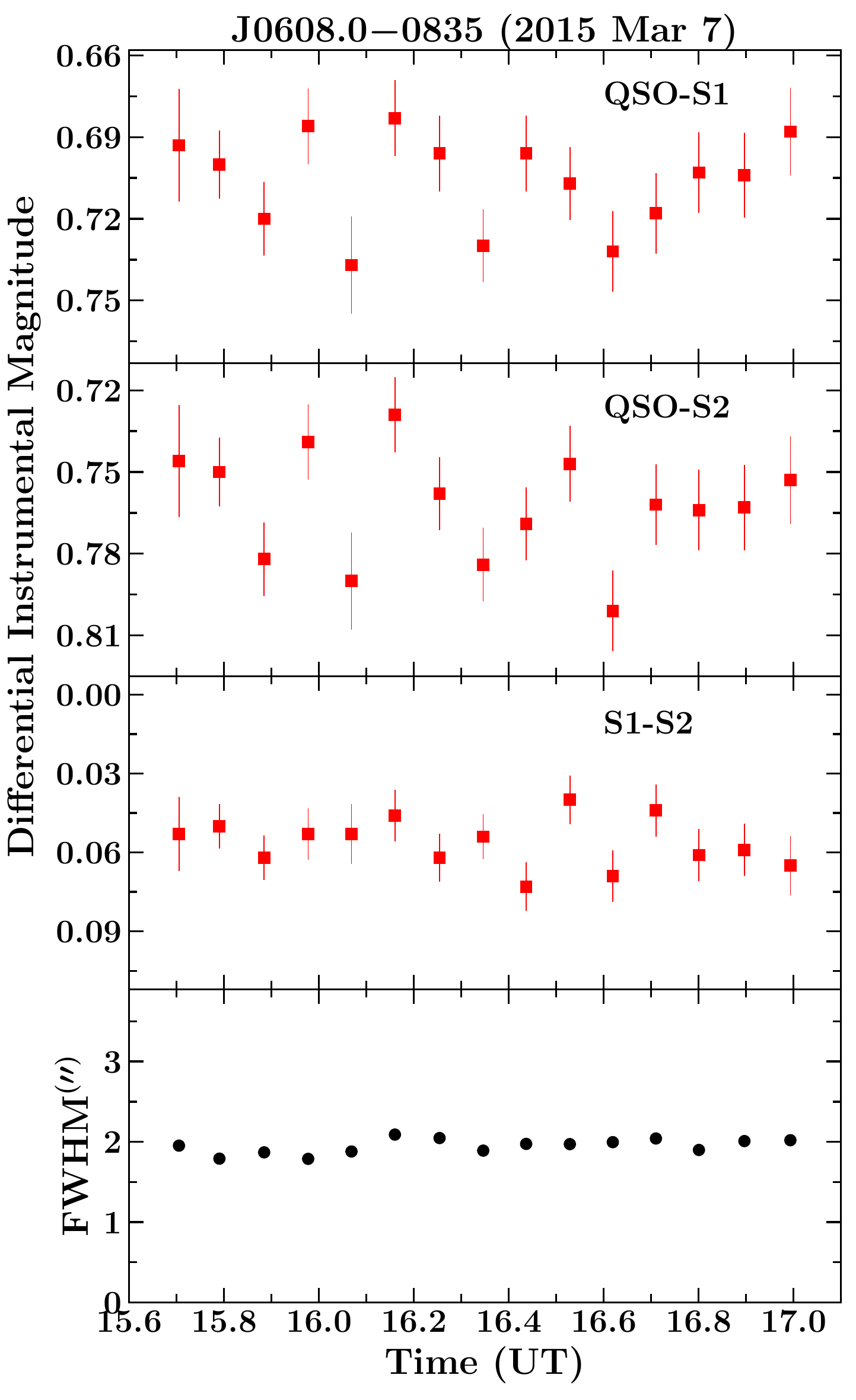}
\includegraphics[width=5.5cm]{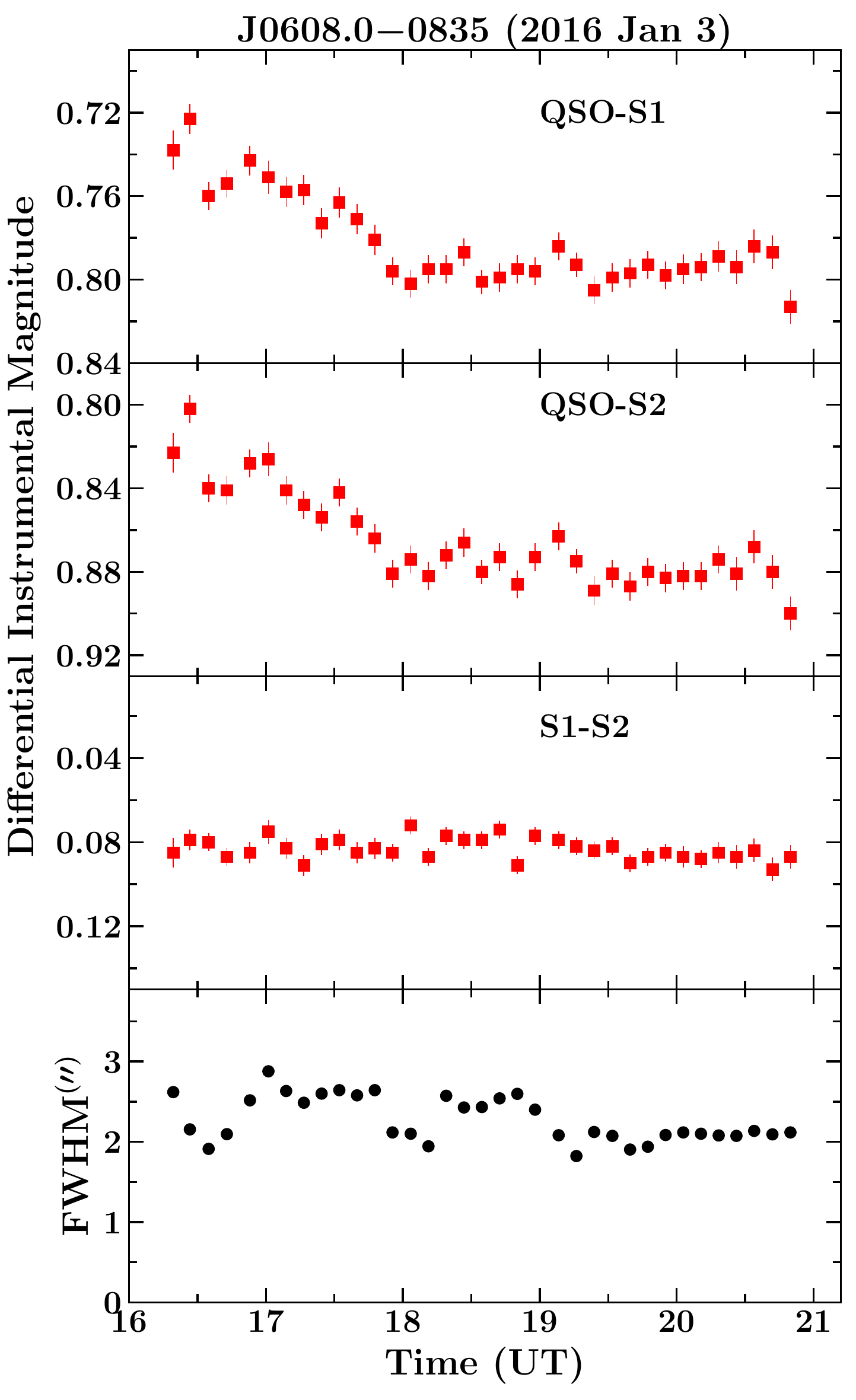}
\includegraphics[width=5.5cm]{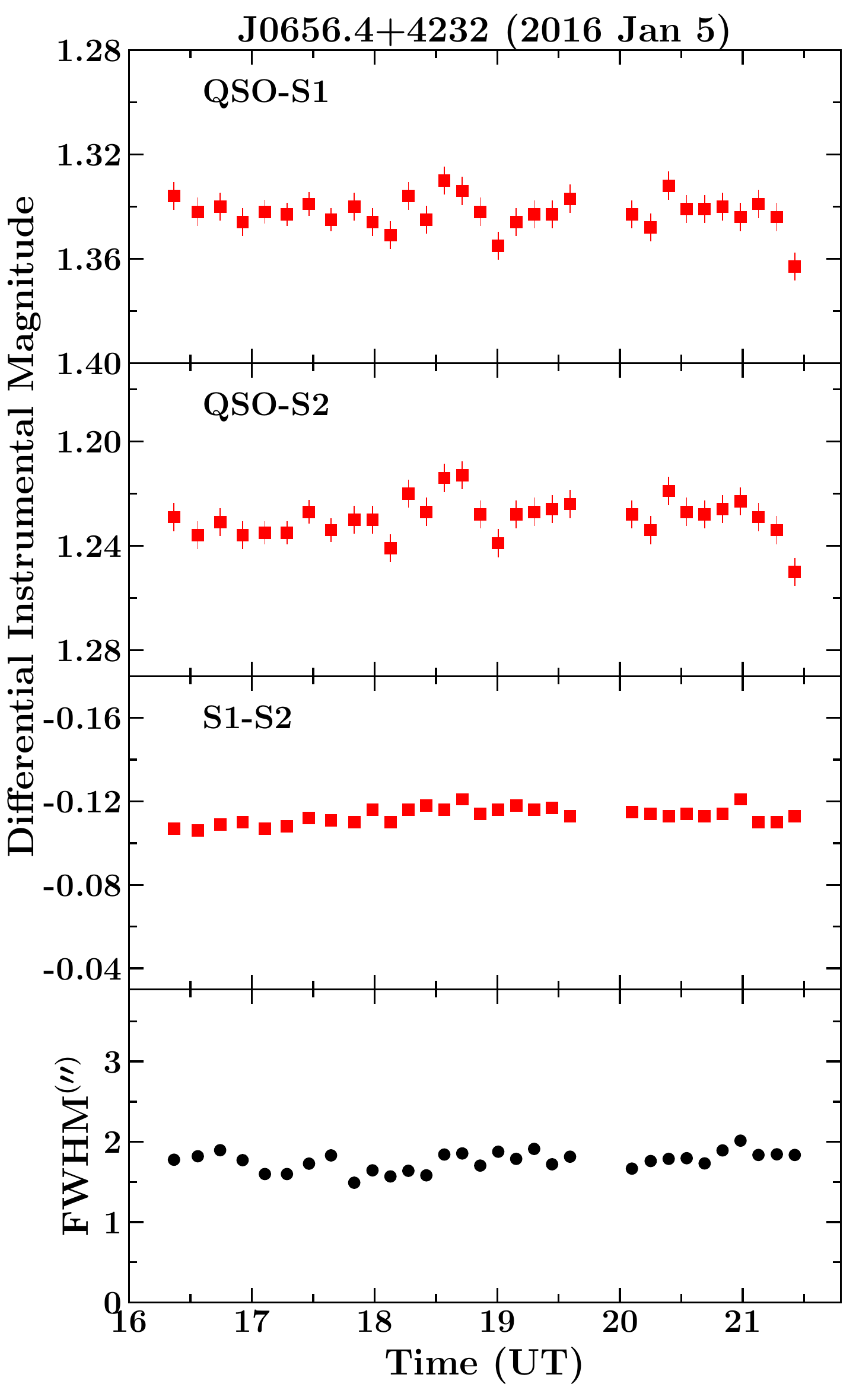}
}
\hbox{
\includegraphics[width=5.5cm]{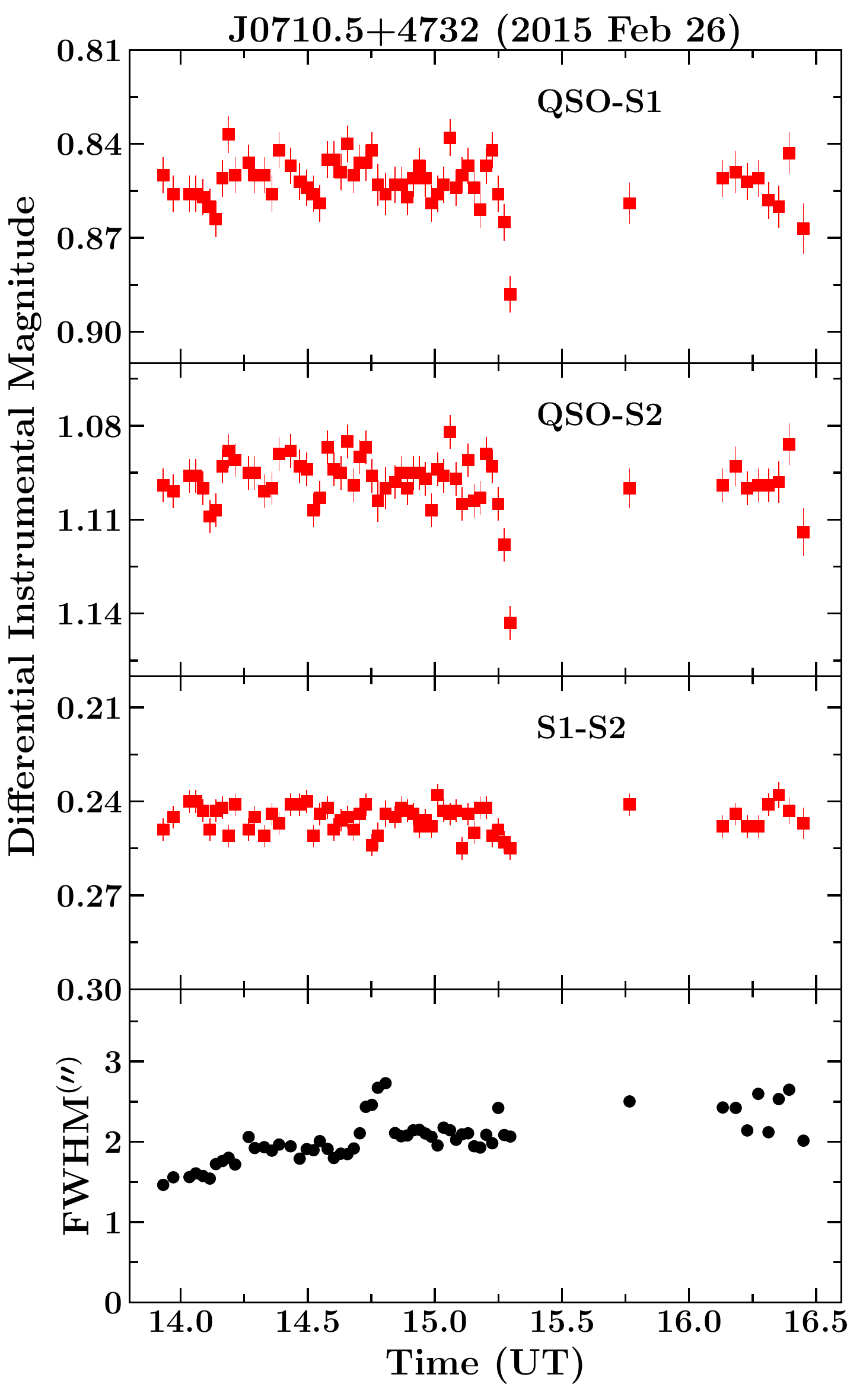}
\includegraphics[width=5.5cm]{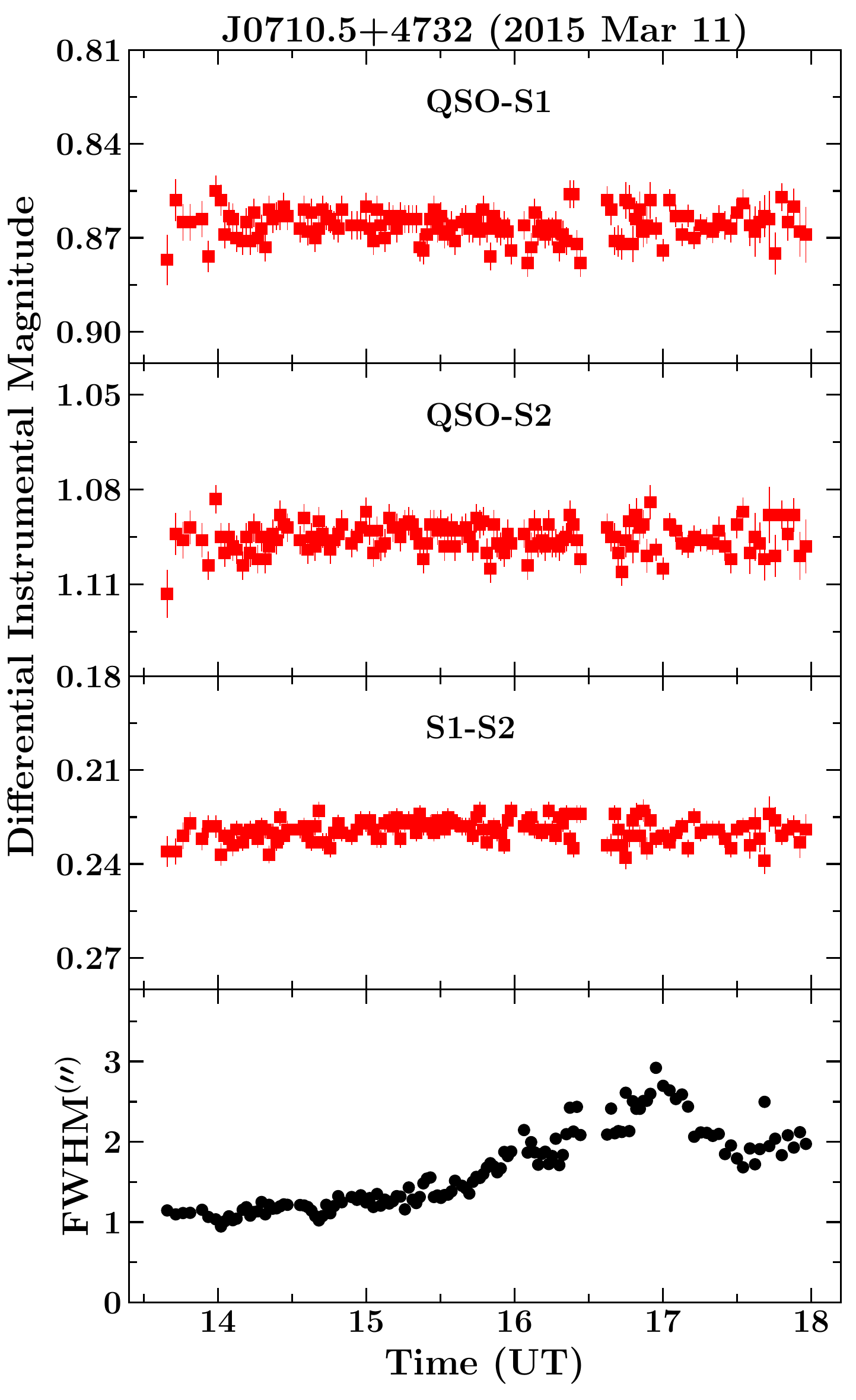}
\includegraphics[width=5.5cm]{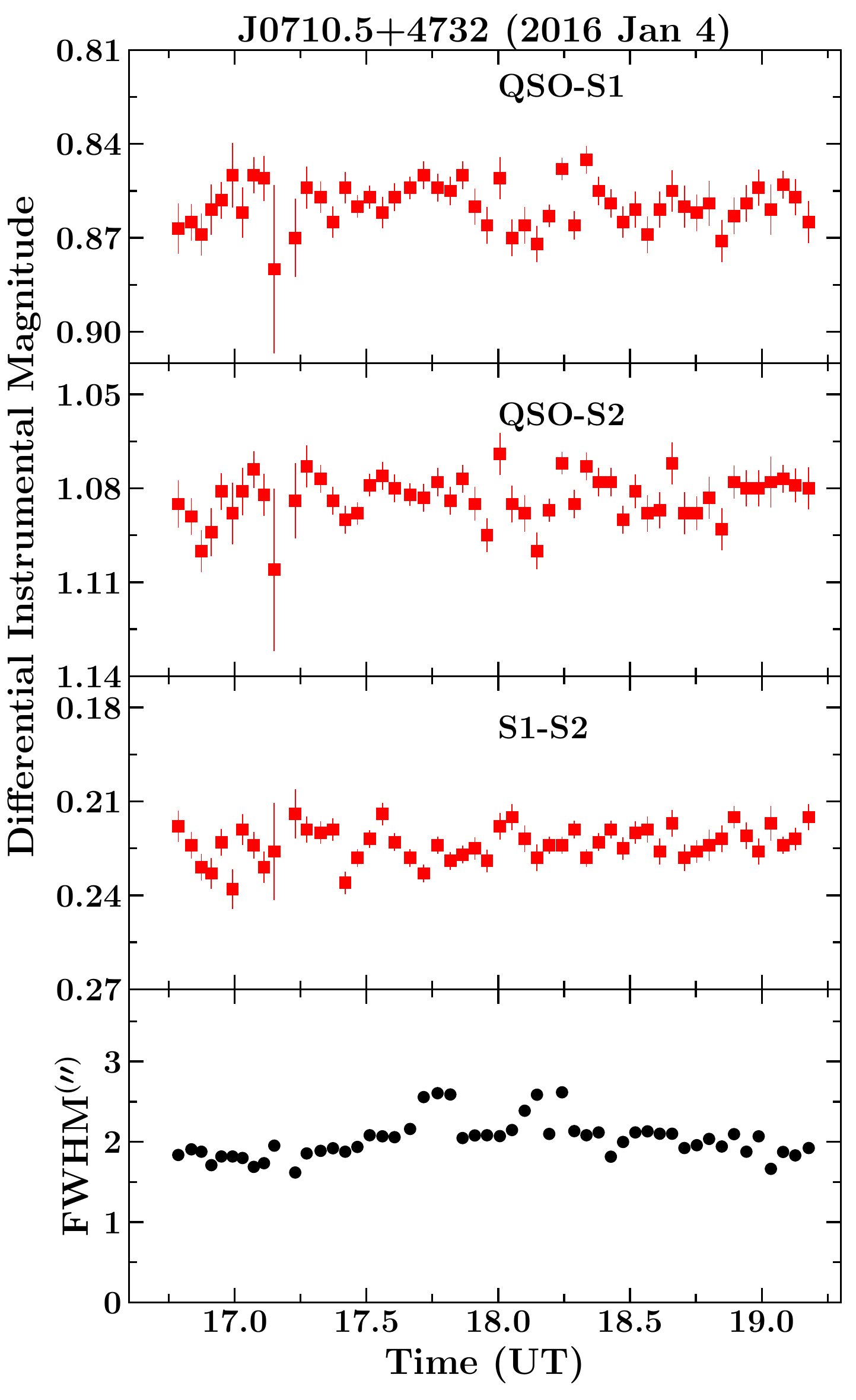}
}
\caption{Intra-night differential light curves of {\it Fermi} blazars monitored during our first observing run. In the bottom panel of each plot is given the variations of the FWHM of the stellar images during the night. QSO refers to the source of interest, while S1 and S2 denote two comparison stars selected to generate the differential light curves.\label{fig_inov1}} 
\end{figure*}

\begin{figure*}
\hbox{
\includegraphics[width=5.5cm]{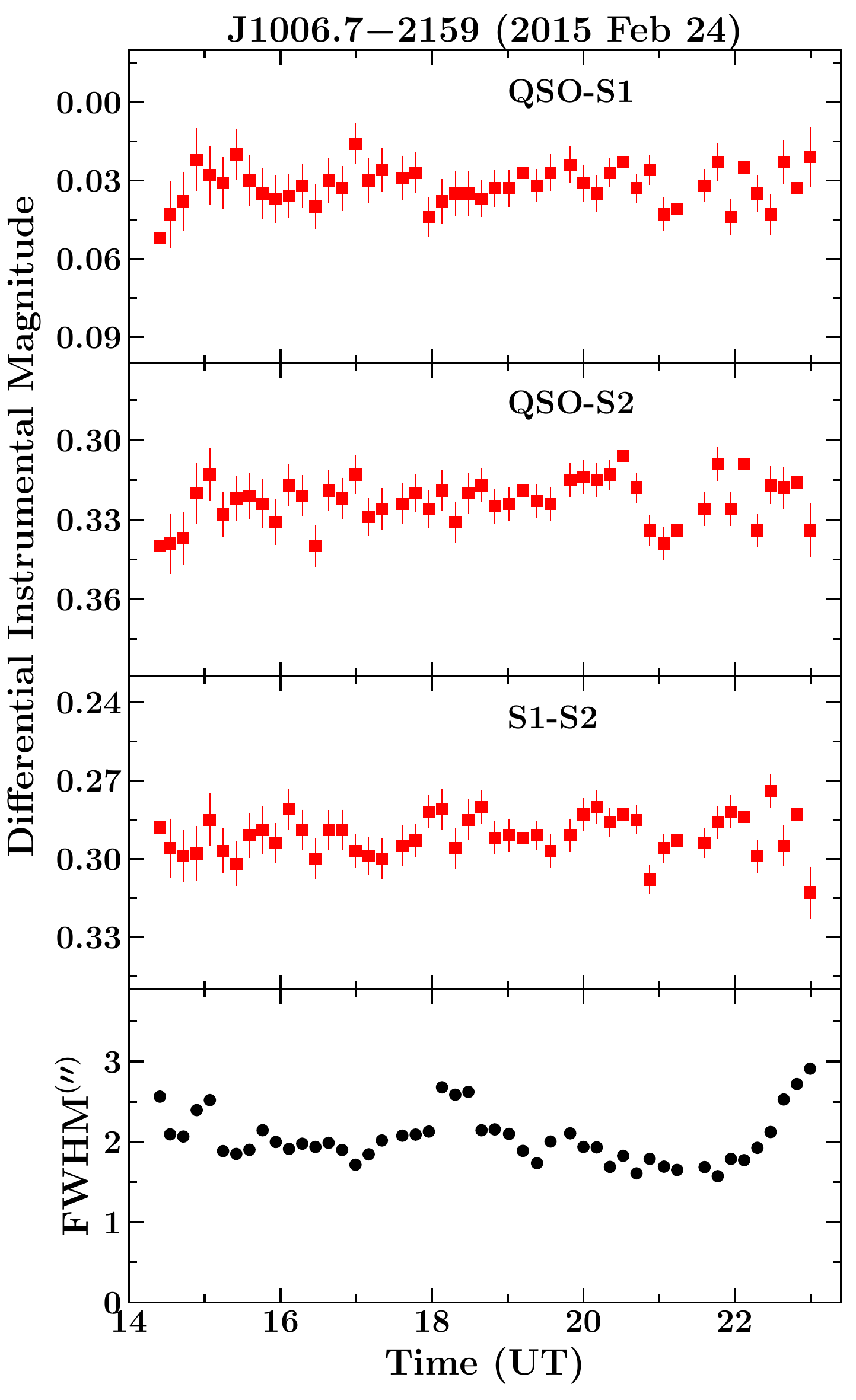}
\includegraphics[width=5.5cm]{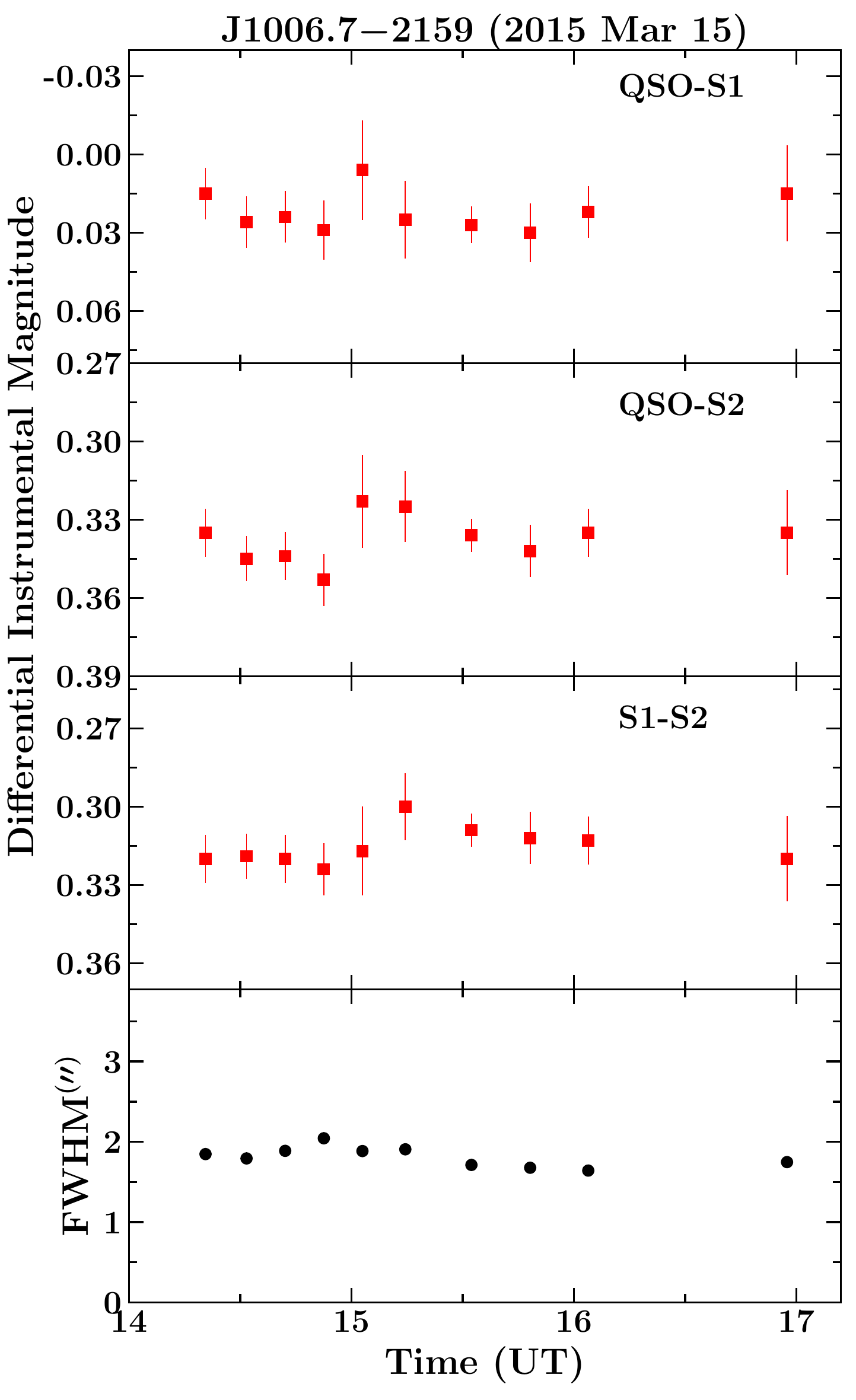}
\includegraphics[width=5.5cm]{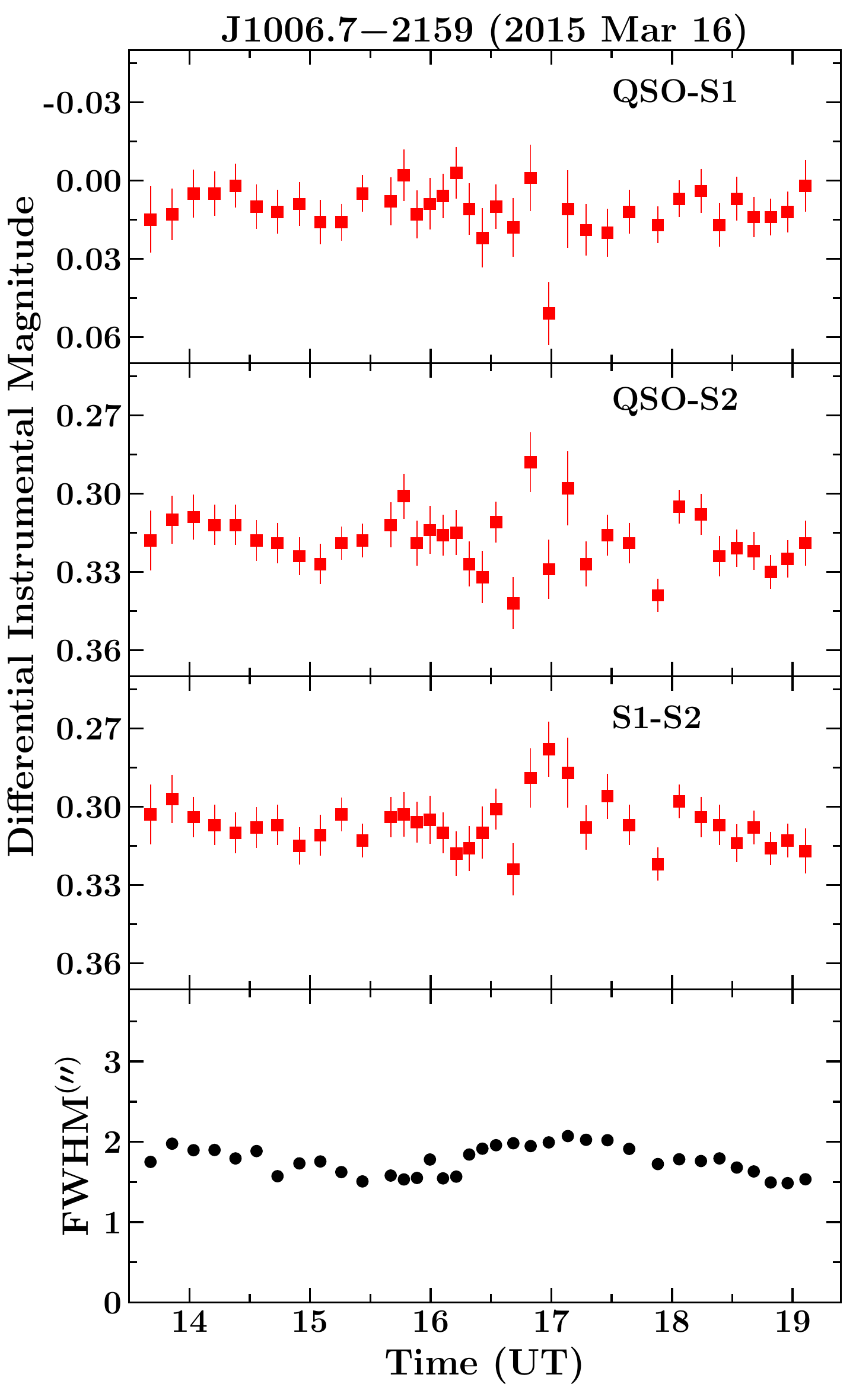}
}
\hbox{
\includegraphics[width=5.5cm]{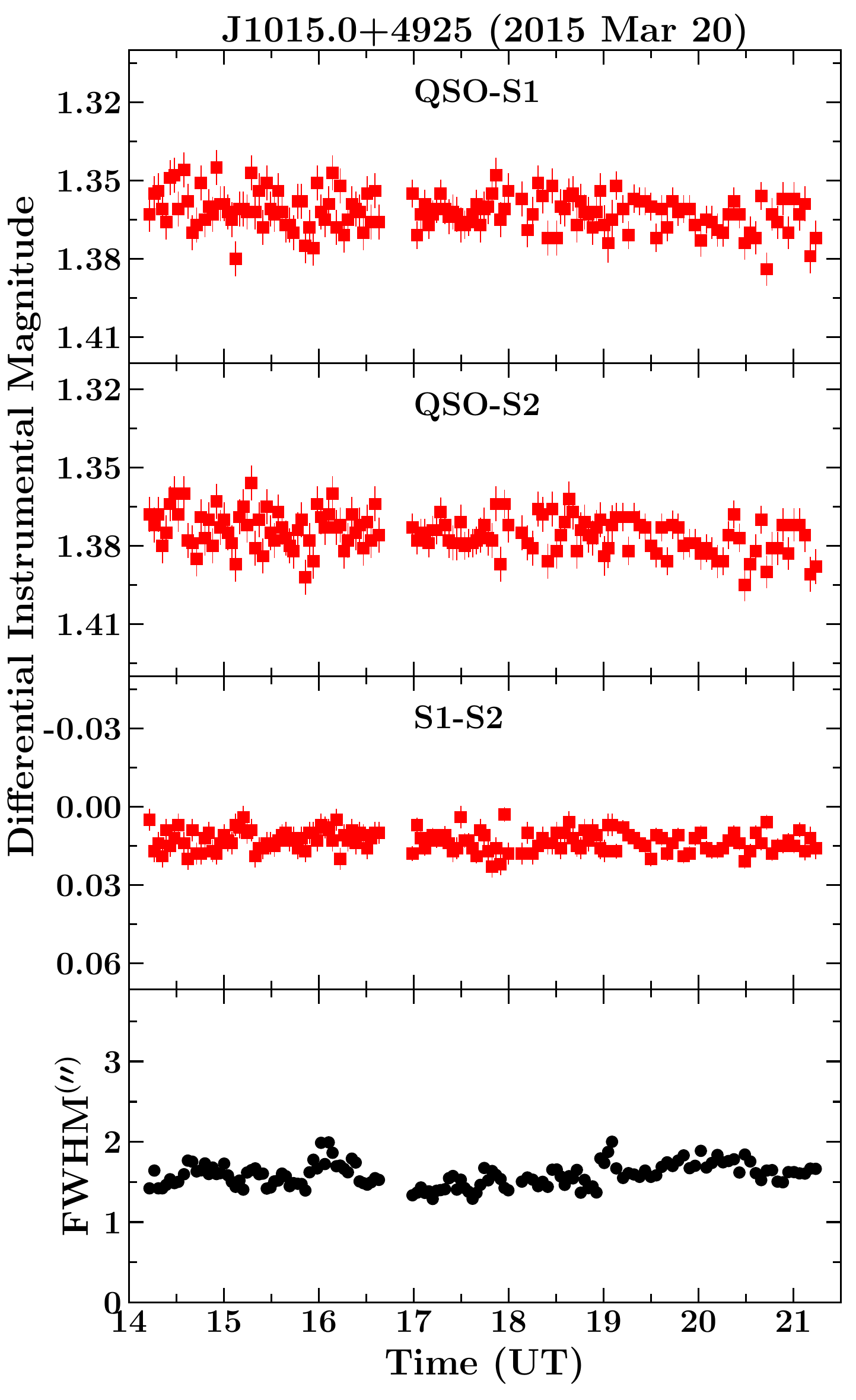}
\includegraphics[width=5.5cm]{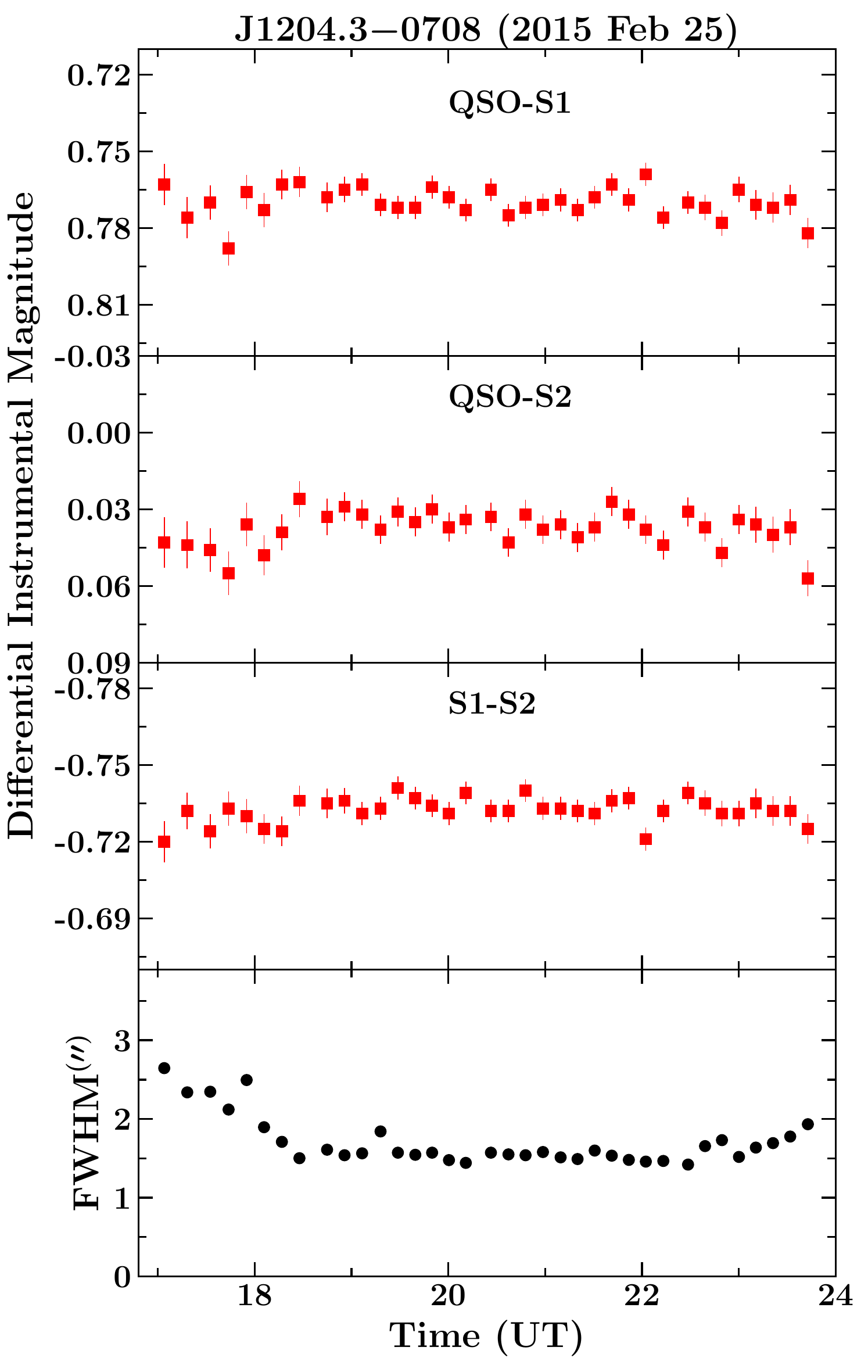}
\includegraphics[width=5.5cm]{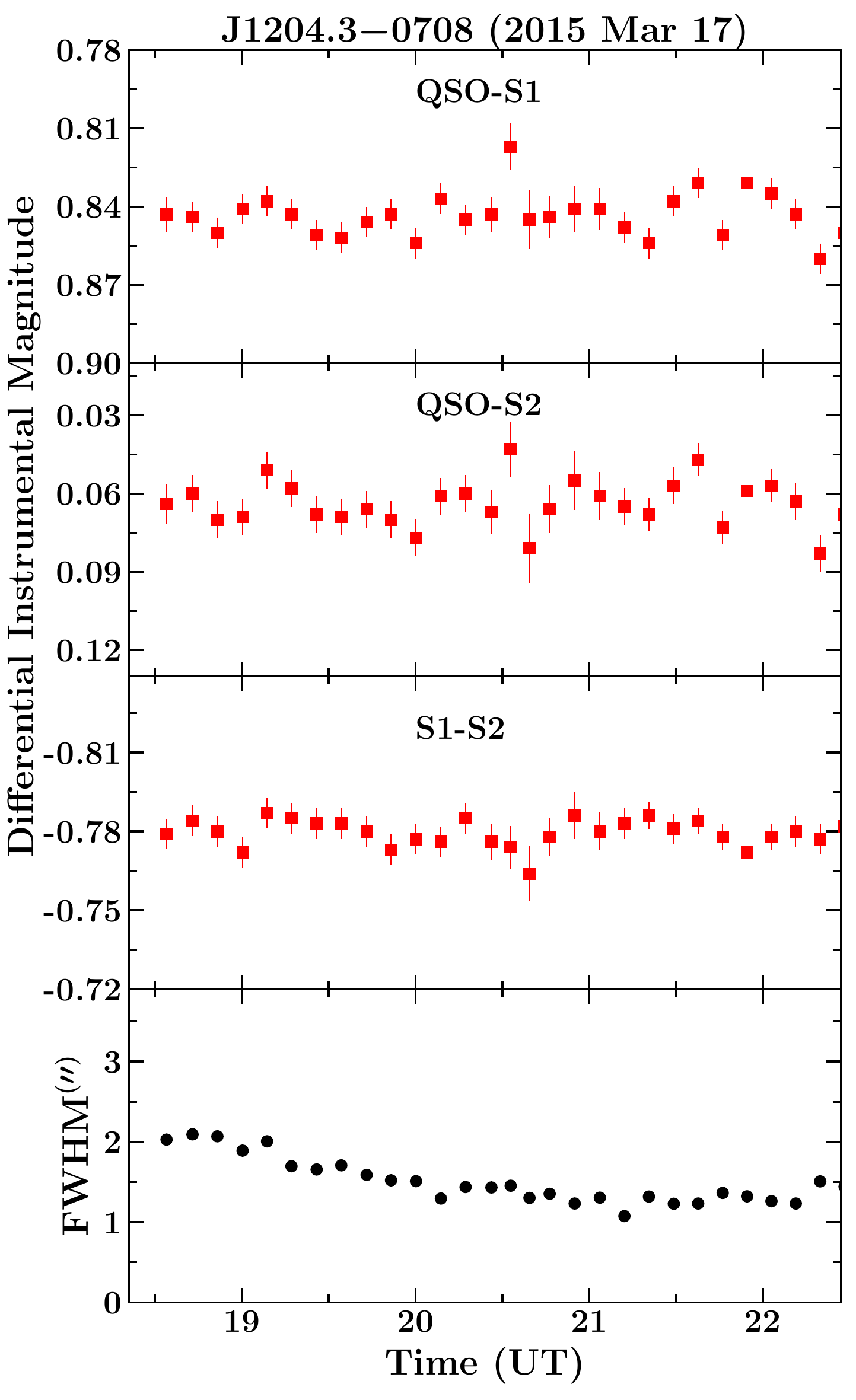}
}
\caption{Intra-night differential light curves of {\it Fermi} blazars. Other information are same as in Figure \ref{fig_inov1}.\label{fig_inov4}} 
\end{figure*}

\begin{figure*}
\hbox{
\includegraphics[width=5.5cm]{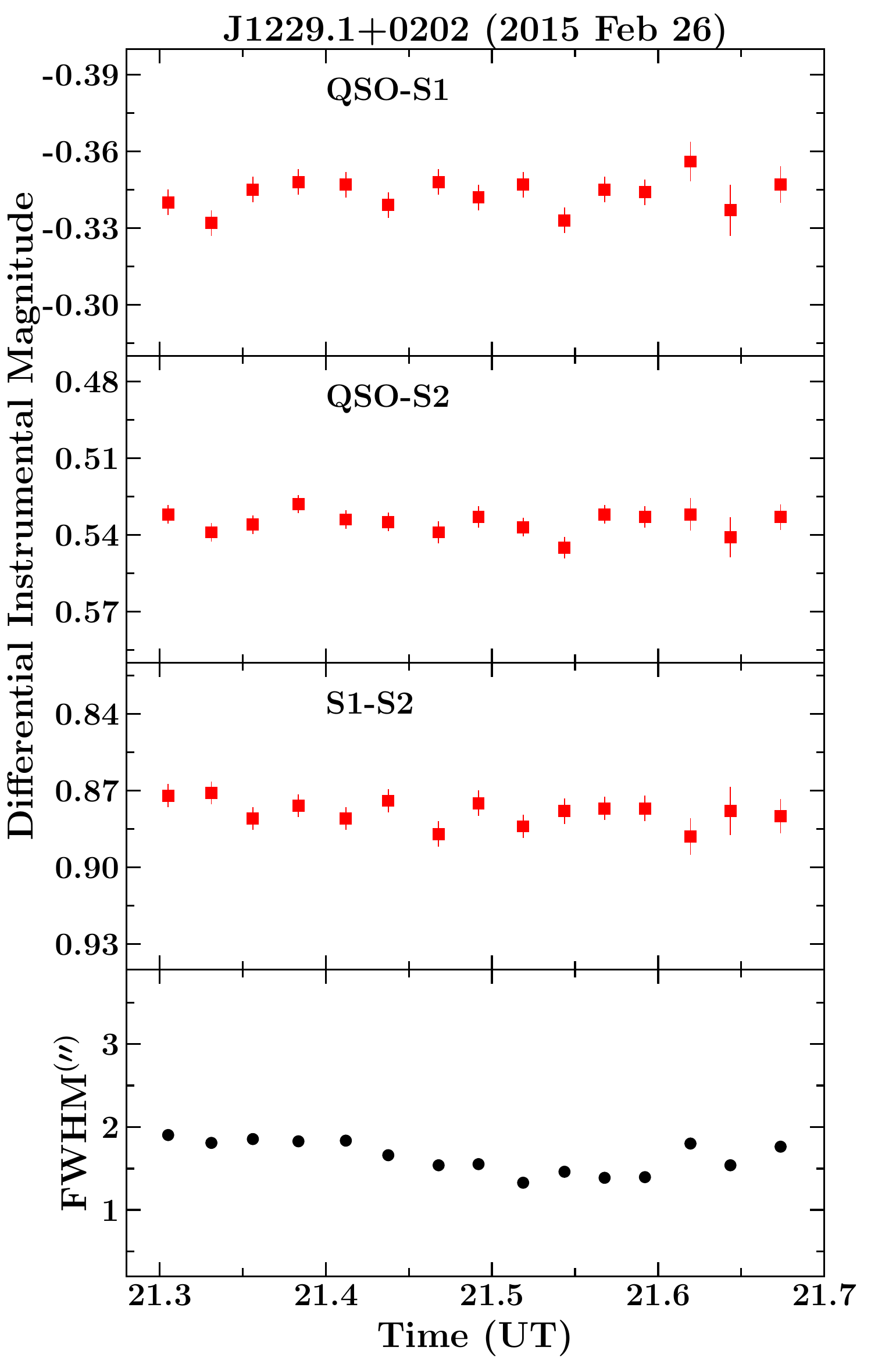}
\includegraphics[width=5.5cm]{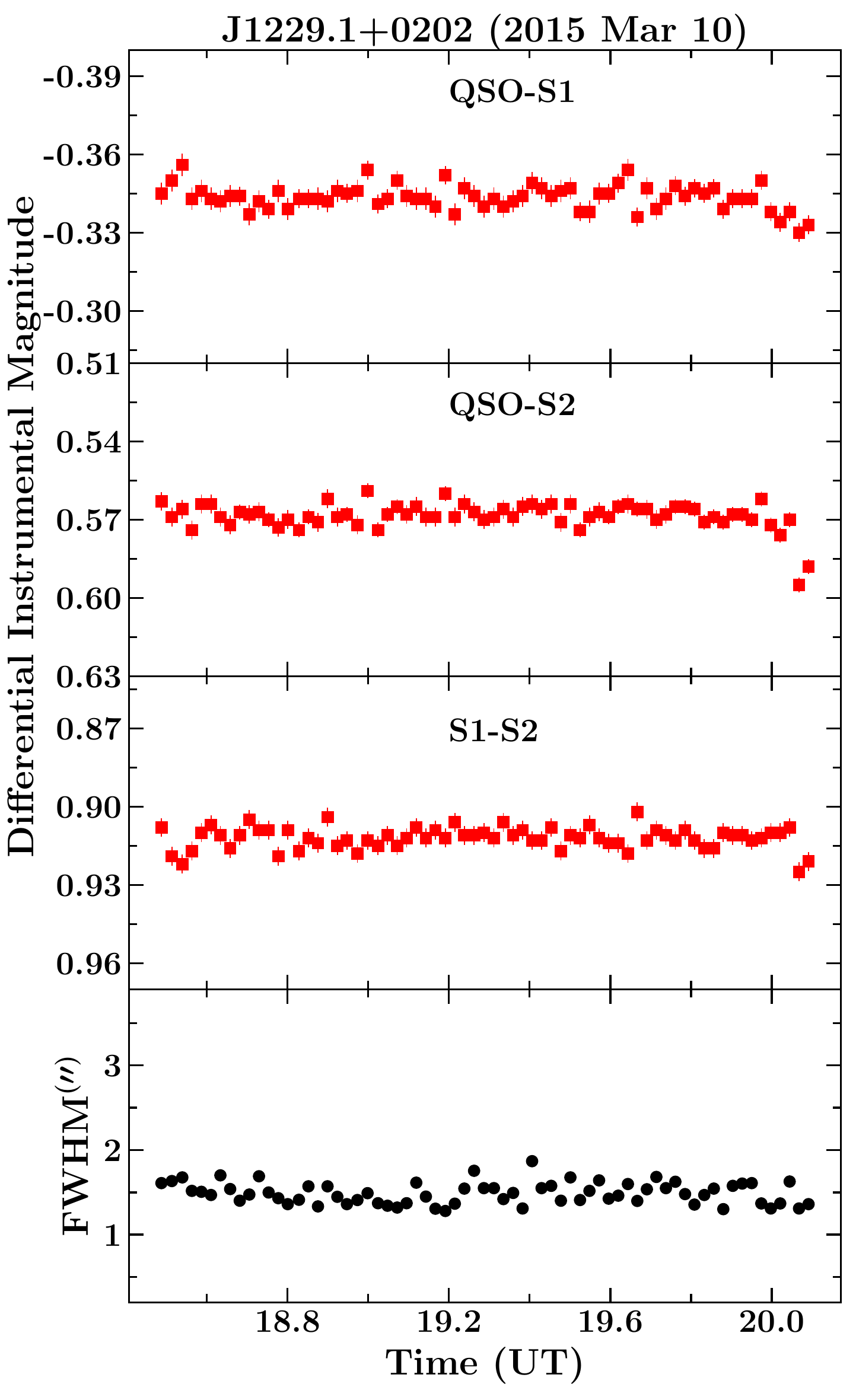}
\includegraphics[width=5.5cm]{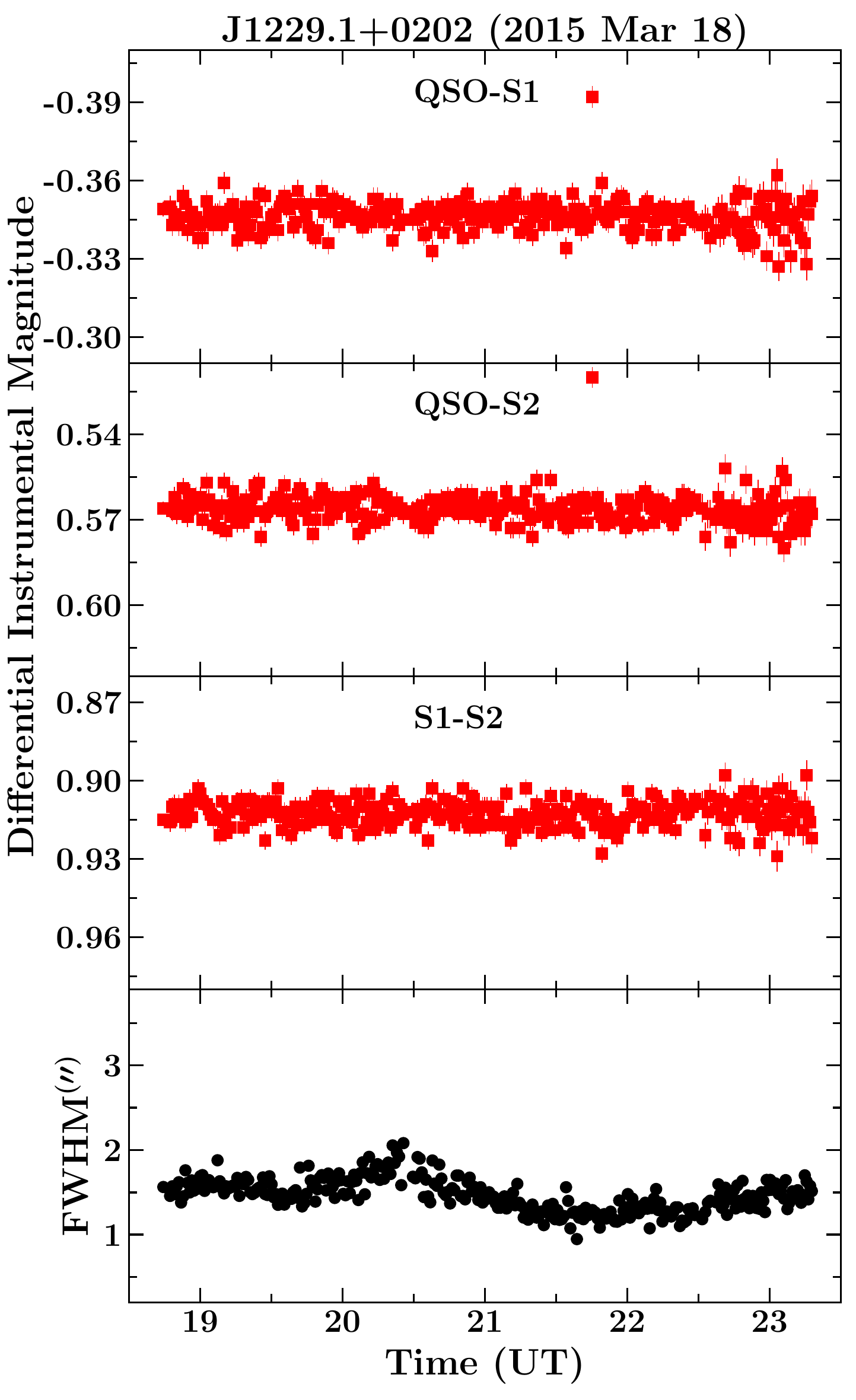}
}
\hbox{
\includegraphics[width=5.5cm]{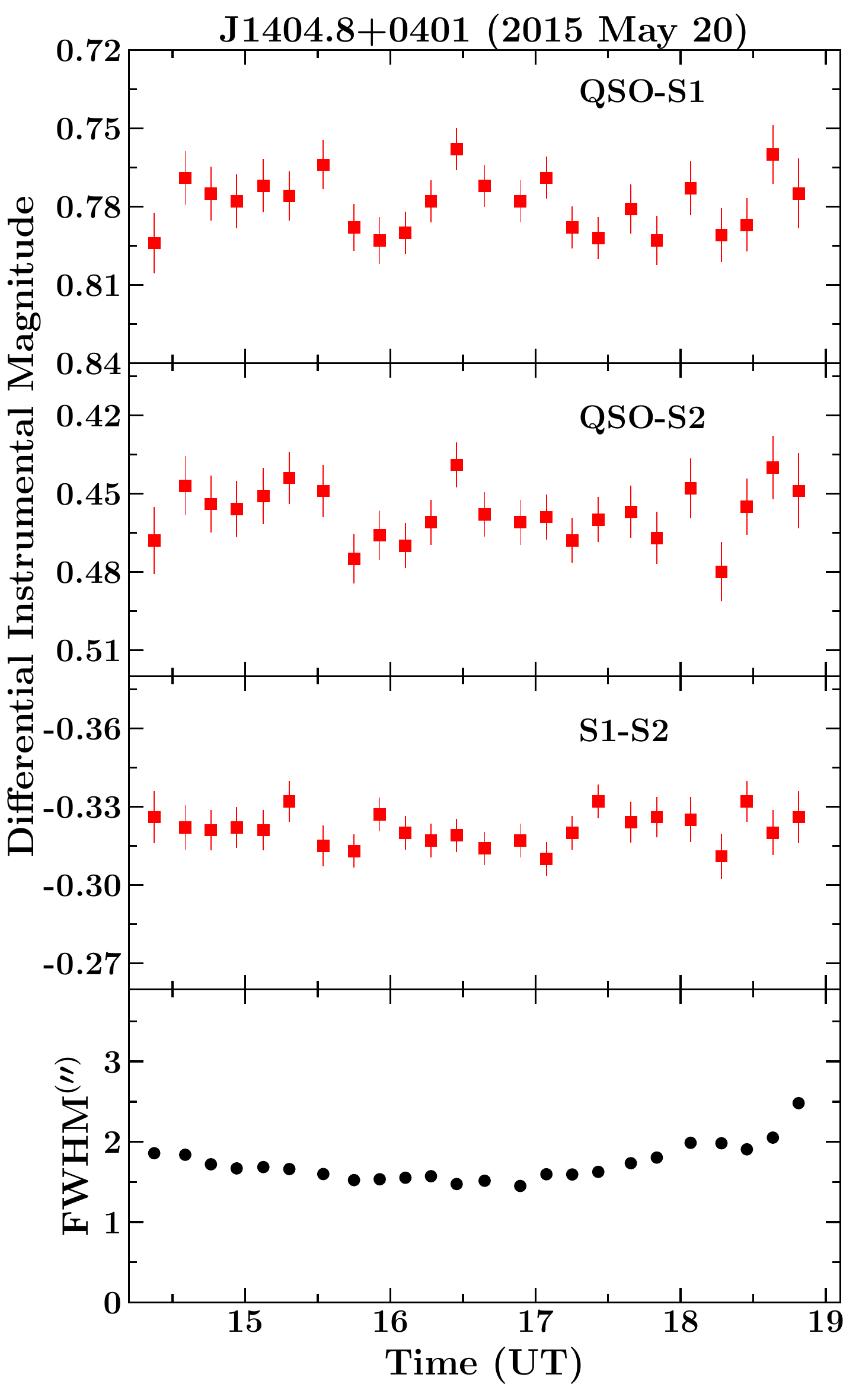}
\includegraphics[width=5.5cm]{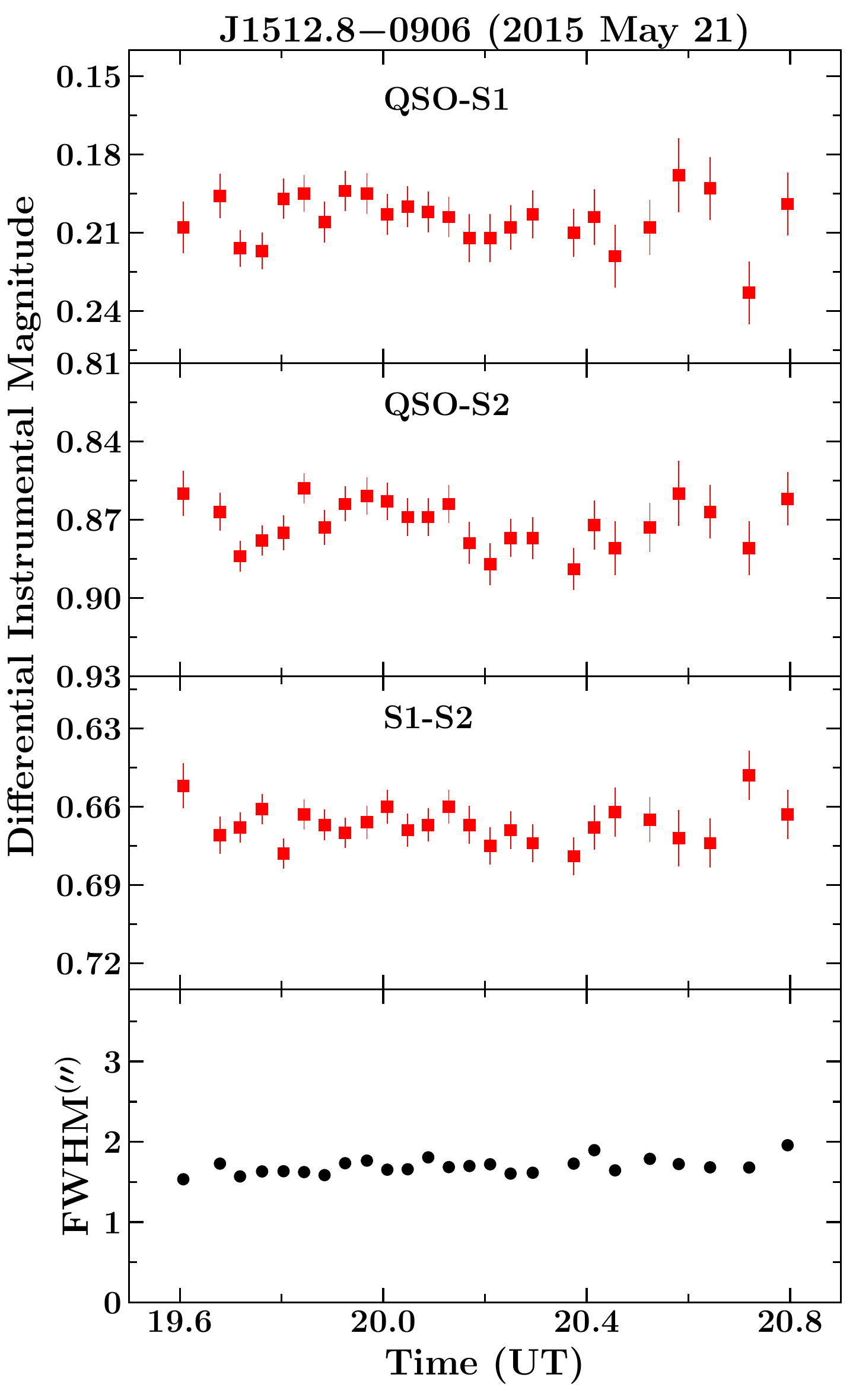}
\includegraphics[width=5.5cm]{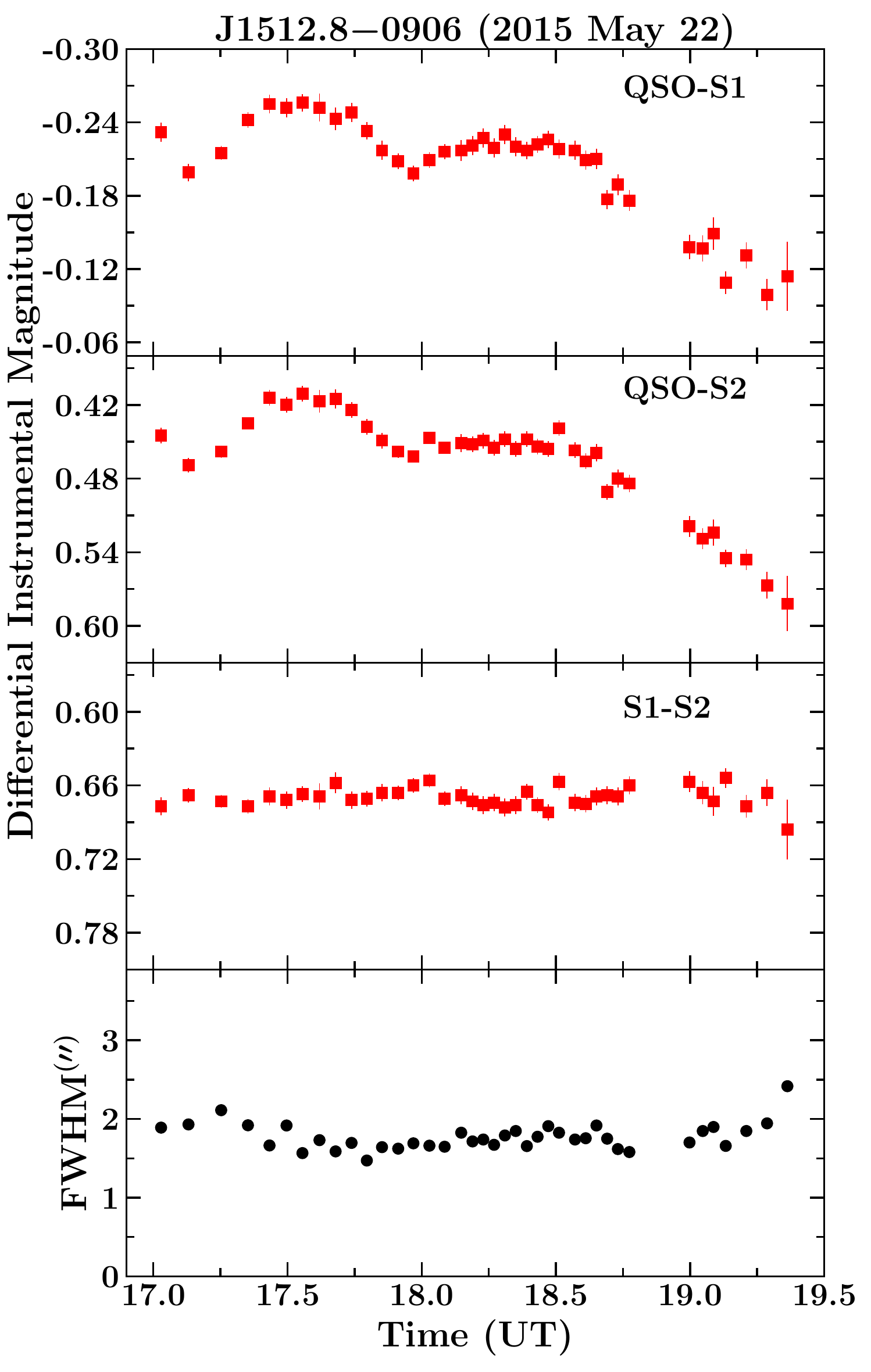}
}
\caption{Intra-night differential light curves of {\it Fermi} blazars. Other information are same as in Figure \ref{fig_inov1}.\label{fig_inov5}} 
\end{figure*}

\begin{figure*}
\hbox{\hspace{2.5cm}
\includegraphics[width=5.5cm]{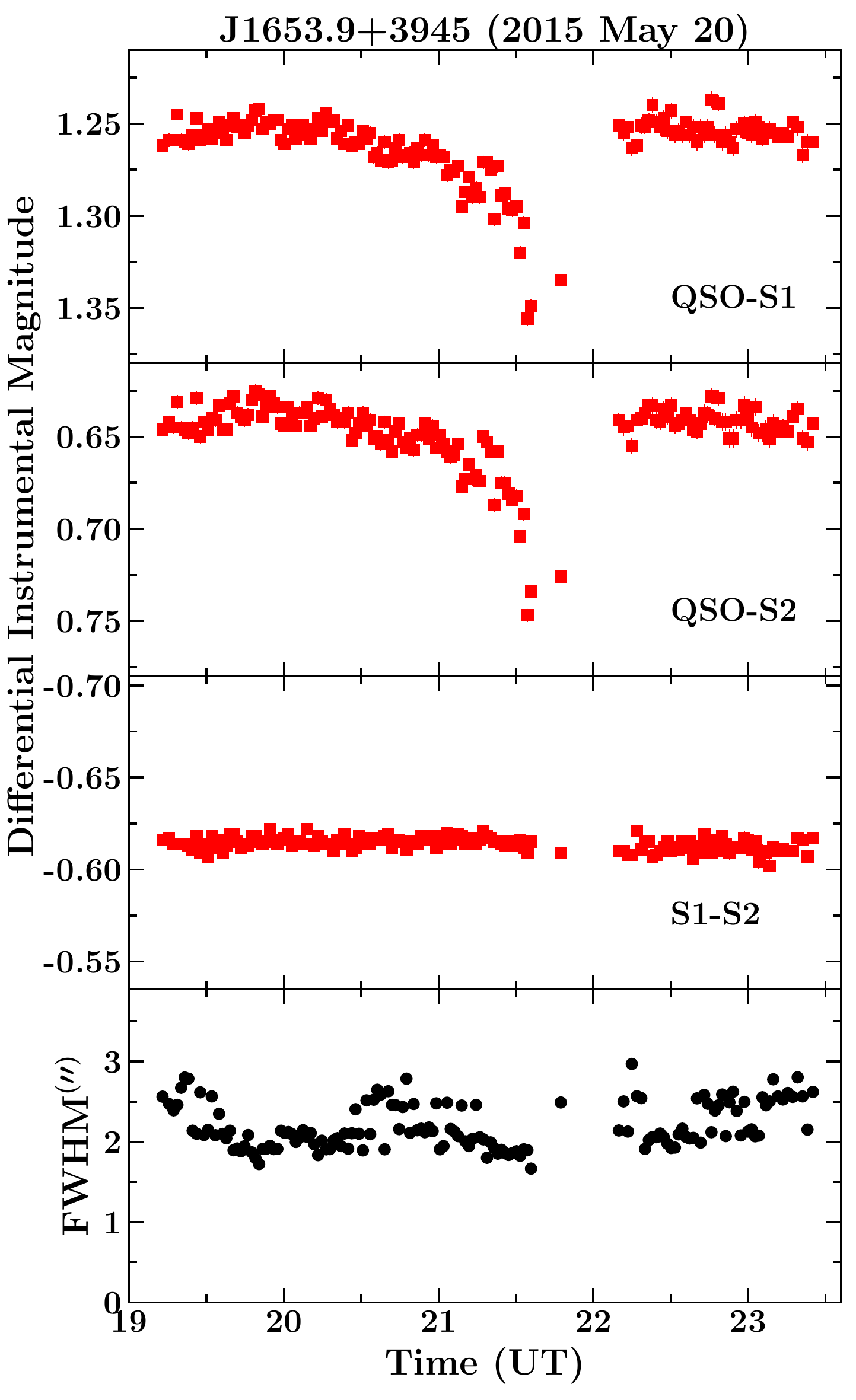}
\includegraphics[width=5.5cm]{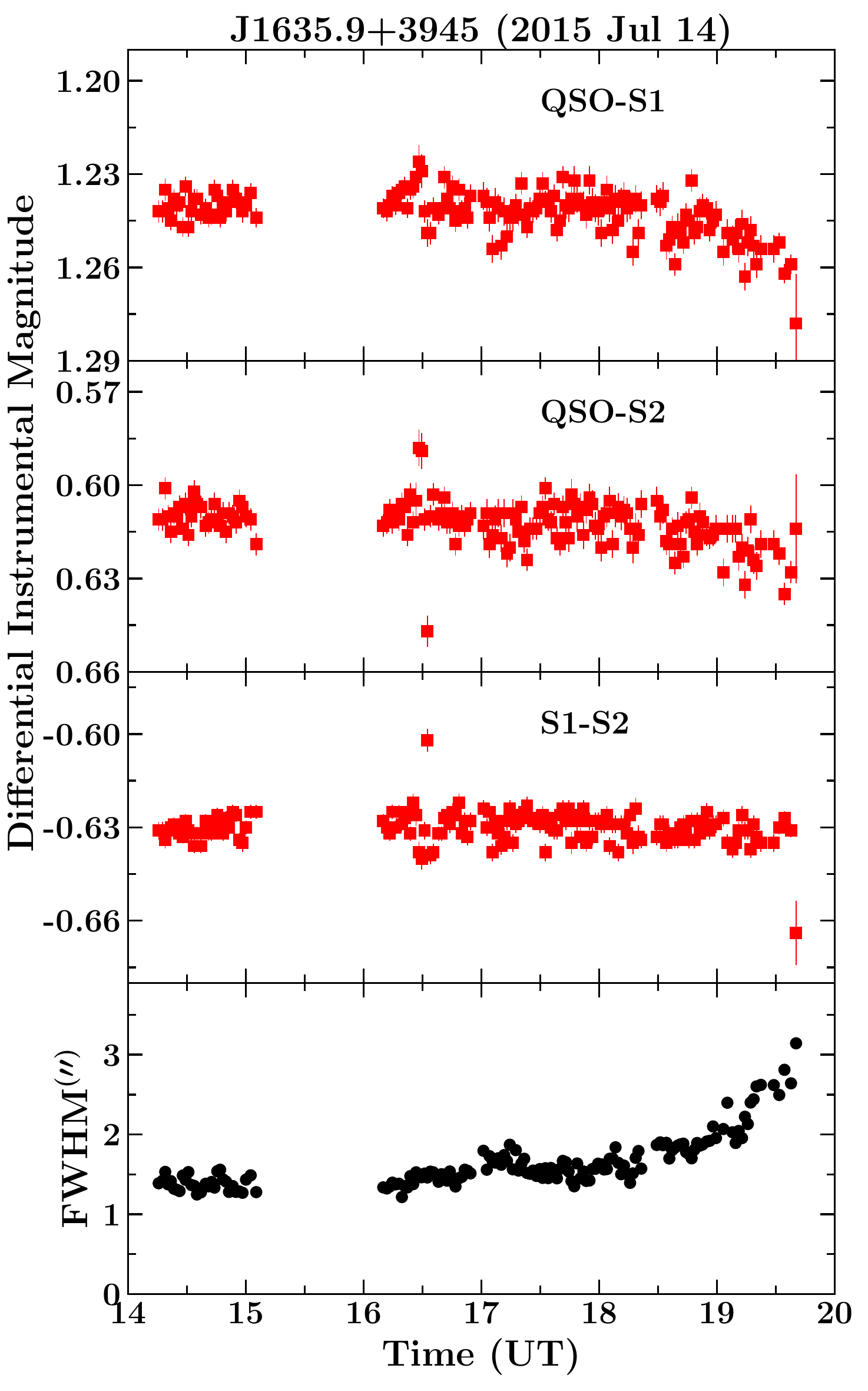}
}
\caption{Intra-night differential light curves of {\it Fermi} blazars. Other information are same as in Figure \ref{fig_inov1}.\label{fig_inov6}} 
\end{figure*}

{\label{sec:results}}
We generate DLCs from the photometric instrumental magnitudes of the target source and a pair of the steady comparison stars. The star-star DLC represent the observational uncertainties and the intrinsic variability of the stars, whereas, quasar-stars DLCs indicate the intrinsic variability of the blazar with respect to steady stars.  A blazar is considered to be variable when it shows correlated flux variations both in time and amplitude relative to the pair of comparison stars. On the other hand, the source is considered as non-variable if it does not show flux variability with respect to both the comparison stars. It should be noted that the uncertainty in the comparison stars magnitudes, as given in Table \ref{tab:com_star}, will not have any effect on the DLCs as we look for the variability in the differential instrumental magnitudes between the source of interest and the comparison stars. The derived results are presented in Figure \ref{fig_inov1}-\ref{fig_inov6}. To statistically confirm the variability, we employ the two criteria, $C$ and $F$ statistics.

\subsection{C-Statistics}
$C$ parameter is the commonly used statistical tool to decide the INOV nature of the sources \citep[][]{1997AJ....114..565J}. It is defined as follows
 \begin{equation}
 \label{eq:c-stat} 
C_1 = \frac{SD_{\rm qso-s1}}{SD_{\rm s1-s2}},~ \\
C_2= \frac{SD_{\rm qso-s2}}{SD_{\rm s1-s2}},
\end{equation}
where $SD_{\rm qso-s_i}$ is the standard deviation of the DLC of the target quasar and the $i^{\rm th}$ comparison star and $SD_{\rm s1-s2}$ is the standard deviation of the star-star DLC. Following \citet[][]{1997AJ....114..565J}, we consider the source to be variable if both $C_1$ and $C_2$ are $\geq2.576$, which corresponds to the 99\% confidence level.

\subsection{Scaled F-statistics}
Another statistical tool to quantify the INOV behavior is the $F$-test and this has been shown to be a more reliable and robust method to characterize the micro-variability \citep[][]{2010AJ....139.1269D}. The $F$ parameter is calculated as
\begin{equation}\label{eq:f-test}
F_1 = \frac {V_{\rm qso-s1}}{V_{\rm s1-s2}},~\\
F_2 = \frac {V_{\rm qso-s2}}{V_{\rm s1-s2}}
\end{equation}
where $V_{\rm qso-s}$ and $V_{\rm s1-s2}$ are the variances of the quasar-star and star-star DLCs. These $F$ values are compared with the critical $F$ value, $F^{\alpha}_{\nu}$, where $\alpha$ is the significance level (taken as 0.01 corresponding to a confidence level of $>99$\%) and $\nu$ (=$N_{\rm p}-1$) is the degree of freedom for the DLC. A source is considered to be variable if both $F$ values are found to exceed $F^{\alpha}_{\nu}$.
 
It is noted in various recent studies \citep[e.g.,][]{2011MNRAS.412.2717J,2014AJ....148...93D} that the $F$ test does not estimate the micro-variability adequately if the comparison stars have brightness significantly different from the target quasar \citep[see also][]{2007MNRAS.374..357C,2015AJ....150...44D}. To compensate for the differences, we need to scale the star-star variance by a factor $\zeta$ \citep[][]{2011MNRAS.412.2717J,2014AJ....148...93D}. Since our analysis procedure is similar to that of \citet[][]{2011MNRAS.412.2717J}, we follow their scaling factor to take into account the brightness differences of the comparison stars. It is defined as follows

\begin{equation}
\zeta=\left[\displaystyle{\frac{\sum_\mathbf{i=1}^{N_{\rm p}}\sigma^2_{i, {\rm err}}({\rm qso-s})/N_{\rm p}}{\sum_{i=1}^{N_{\rm p}}\sigma^2_{i, {\rm err}}({\rm s1-s2})/N_{\rm p}}}\right] \equiv \frac{\langle\sigma^2({\rm qso-s})\rangle}{\langle\sigma^2(s{\rm 1-s2})\rangle},
\label{eq:omega}
\end{equation}
where $\sigma^2_{\rm i,err}({\rm qso-s})$ and $\sigma^2_{\rm i,err}({\rm s1-s2})$, respectively, are the errors on individual points of the quasar-star and star-star DLCs. With this, the scaled F-value, $F^{s}$, can be computed as,

\begin{equation}
\label{eq.fstest}
 F_{1}^{s}=\frac{V_{\rm qso-s1}}{\zeta~V_{\rm s1-s2}},~ \\
 F_{2}^{s}=\frac{V_{\rm qso-s2}}{\zeta~V_{\rm s1-s2}}.
\end{equation}
In this work, we use this scaled $F$ test to ascertain the INOV nature of the target blazars.

It has been reported in recent works that claiming the presence of the INOV using $C$ statistics is not correct as it is almost too conservative \citep[e.g.,][]{2010AJ....139.1269D,2014AJ....148...93D}. However, it is also important to consider the fact that this method might be a more compelling measure of the INOV, especially when the comparison stars are not steady \citep[][]{2017MNRAS.tmp...73Z}. We also adopt this tool to compare our results with that obtained in earlier INOV studies of blazars where only $C$ statistics was employed \citep[see, e.g.,][]{2001AJ....122.2901D,2004MNRAS.348..176S}. Accordingly, we report the results of both $C$ and scaled $F$-statistics in Table \ref{tab:stat}.

\subsection{Amplitude of Variability}
The INOV variability amplitude ($\psi$) quantifies the actual variation exhibited by the target source on any given night, after correcting for the errors in the measurements. It is defined as follows \citep[][]{1999A&AS..135..477R}

\begin{equation}\label{eq:amplitude} 
\psi = 100 \sqrt{A_{\rm max} - A_{\rm min})^{2} - 2V_{\rm s1-s2}}\%
\end{equation}
with \\
$A_{\rm max}$ = maximum in $\gamma$-NLSy1 differential light curve,
\\ $A_{\rm min}$ = minimum in $\gamma$-NLSy1 differential light curve,
\\ $V_{\rm s1-s2}$ = variance in the star$-$star DLC

The results of this exercise are presented in Table \ref{tab:stat}.

\subsection{Duty Cycle}
Duty cycle (DC) is computed to determine the fraction of the time an object shows the variability. This is because a blazar may not show flux variations on all observing nights. The DC is evaluated by taking the ratio of the time over which the target blazar exhibits variability to the total monitoring time. Following \citet[][]{1999A&AS..135..477R}, it is defined as

\begin{equation}\label{eq:dc}
DC  = 100\frac{{\sum_{k=1}^{N_{\rm p}}} Q_k(1/\Delta t_k)}{\sum_{k=1}^{N_{\rm p}} (1/\Delta t_k)} {\rm ~\%}
\end{equation}
where $\Delta t_{k}$ = $\Delta t_{k, {\rm obs}}(1 + z)^{-1}$ is the redshift corrected time interval of the monitoring session of a source on the $i^{\rm th}$ night. 
$Q_{k}$ is set equal to 1 if the source is found to be variable, otherwise $Q_{k}=0$.

We find a DC of 26\% and 25\% for the entire sample, when considering the $C$-statistics and the scaled $F$-statistics, respectively. It is found to be 14\% according to the $C$-statistics and 16\% following the $F$-test, for FSRQs. For BL Lac objects, DC is derived as 40\% using both the $C$ and the $F$-tests. Alternatively, if we define the variability criteria with the detection of the high amplitude of variability \citep[$\psi>3$\%, e.g.,][]{2004JApA...25....1S}, the overall DC increases to 66\%. Considering FSRQs and BL Lac objects separately, it is estimated as 52\% and 86\%, respectively. These results confirms the earlier findings that BL Lac objects show a larger INOV compared to FSRQs. On dividing the sources according to the location of their $\nu^{\rm peak}_{\rm syn}$, we find that LSP blazars exhibit a substantially larger INOV (DC$\sim$59\%) compared to ISP and HSP sources that have a DC of $\sim$11\% and 13\%, respectively.

\section{Discussion on Individual Sources}{\label{sec:objects}}
Blazars are known to vary at all timescales \citep[e.g.,][]{2011ApJS..194...29R}. If we assume a co-spatial origin for the observed multi-wavelength radiation from blazars, a flaring state in one band should be reflected at other frequencies, though with different timescales due to different scales of the Doppler boosting \citep[][]{1995ApJ...446L..63D}. Since the sources monitored under our observing campaign are known $\gamma$-ray emitters, we expect to observe a high amplitude INOV during the elevated $\gamma$-ray activity state. On the other hand, an optical flux variability without a $\gamma$-ray counterpart indicates the change in the magnetic field that enhances synchrotron emission or/and the location of the emission region being close to the black hole where the jet bulk Lorentz factor is small, as a possible causes of the orphan optical flares \citep[e.g.,][]{2007MNRAS.382L..82G,2013ApJ...763L..11C}. The latter becomes a major factor for FSRQs whose $\gamma$-ray emission is known to be dominated by the external Compton processes \citep[][]{2009ApJ...704...38S}. In other words, it is possible to detect a short term INOV from a non-thermal synchrotron dominated jet, even during the $\gamma$-ray quiescence. 

\begin{figure*}
\centering
\hbox{
\includegraphics[width=9cm]{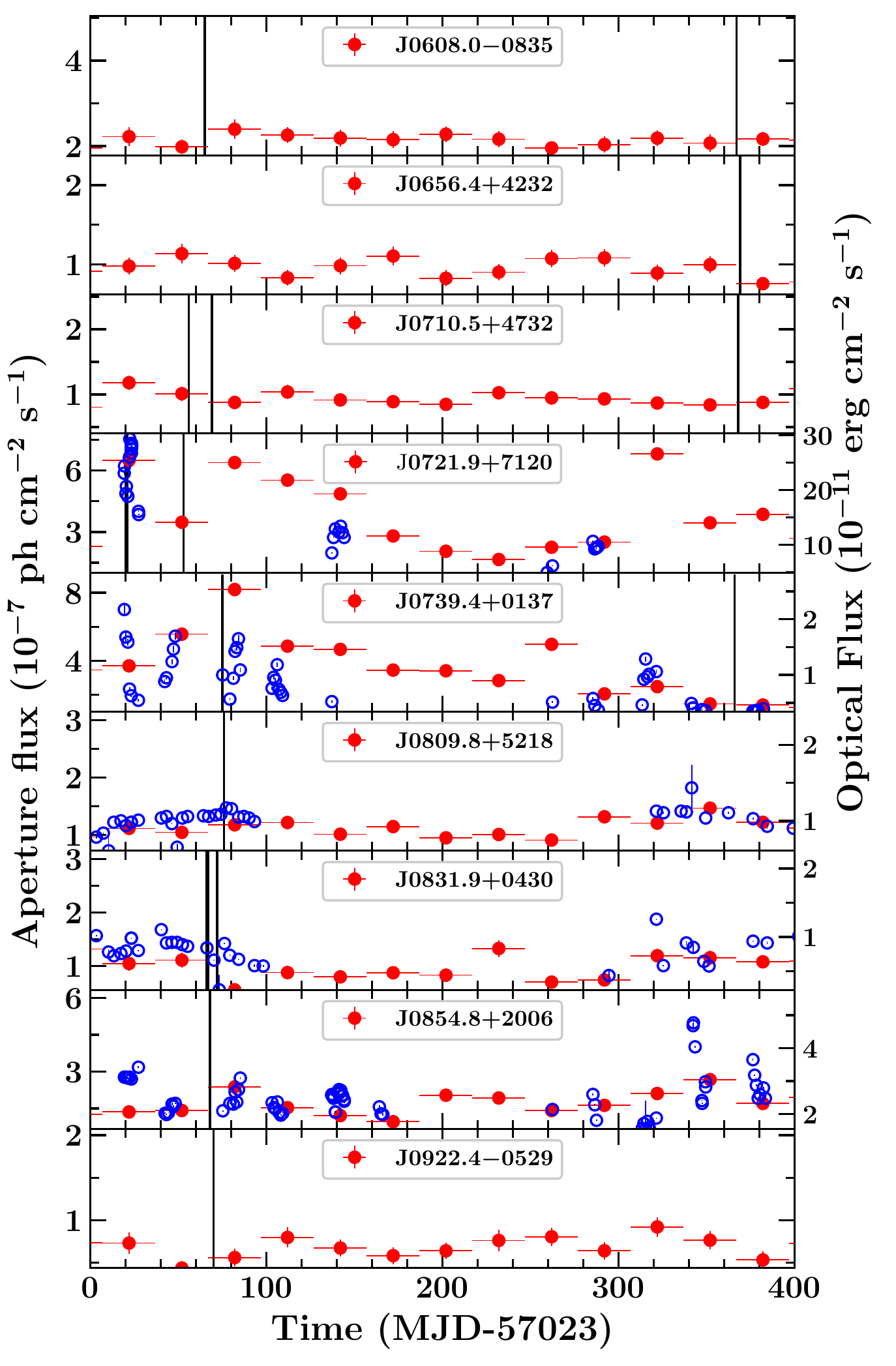}
\includegraphics[width=9cm]{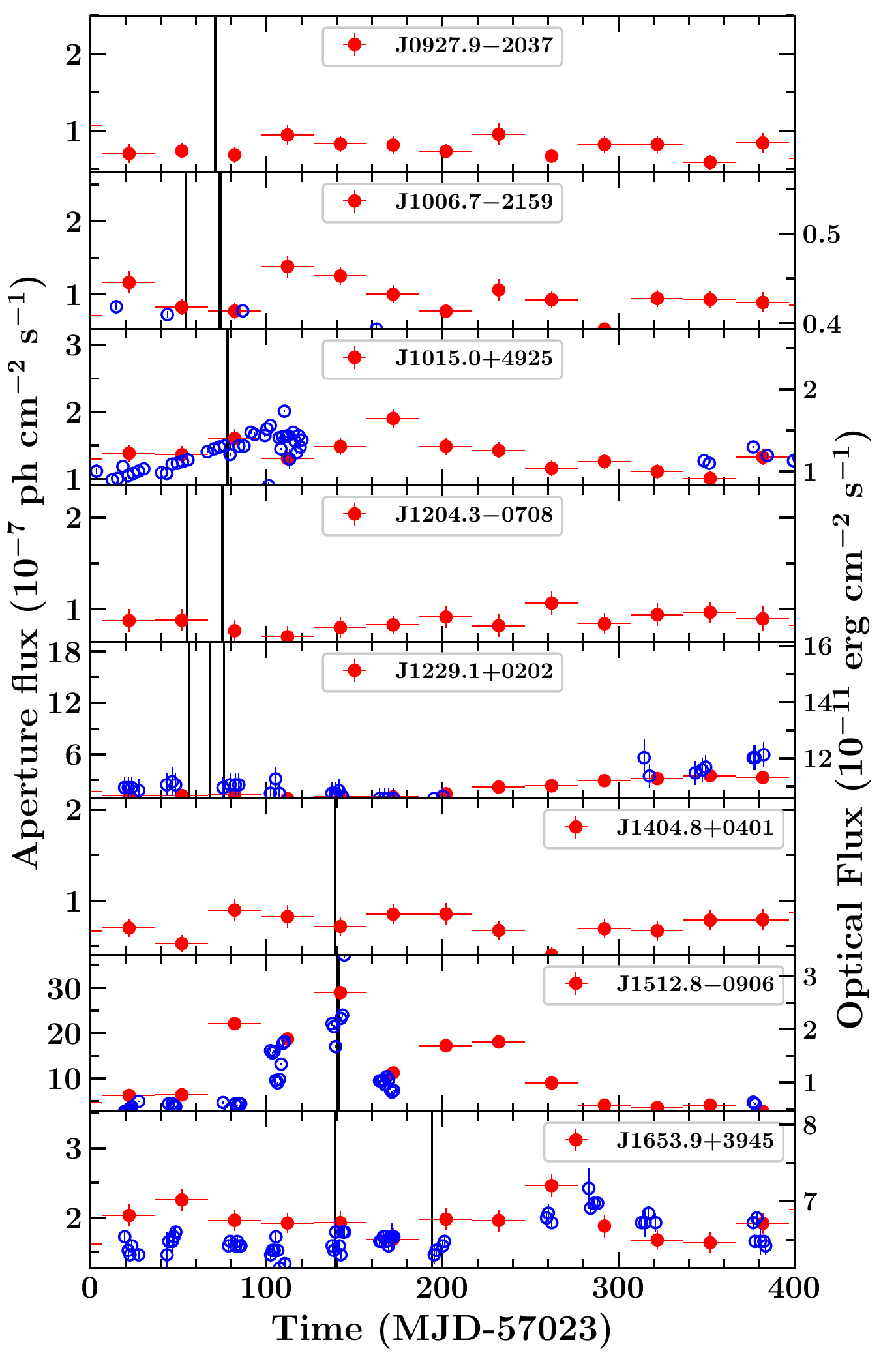}
}
\caption{Monthly binned aperture photometric light curves of \fermi~blazars since 2015 January 1 (MJD 57023). Overplotted are the long term optical observations of 10 objects, shown with blue empty circles, that were carried out at the Steward, KAIT, and SMARTS observatories. The vertical lines denote the epoch of the optical monitoring from the 1.3 m JCBT. Note that the ranges on y-axes are plotted between the minimum and maximum of the flux observed since the beginning of the \fermi~operation to show the relative activity level of the sources during the observing epoches. See the text for details. \label{fig_gamma1}} 
\end{figure*}

To understand the optical-$\gamma$-ray connection in deeper perspective, we plot the publicly available, monthly binned, aperture photometric $\gamma$-ray light curves\footnote{http://fermi.gsfc.nasa.gov/ssc/data/access/lat/4yr\_catalog/ap\_lcs.php} for all 17 blazars in Figure \ref{fig_gamma1}. It should be noted that the $\gamma$-ray aperture photometry may not be suitable for a detailed scientific investigation, however, since our primary purpose is just to get an idea about the $\gamma$-ray activity state during the epoch of the optical monitoring, these results can be used. In Figure \ref{fig_gamma1}, the x-axis covers the time period since 2015 January 1 (MJD 57023) till 2016 February 5 (MJD 57423). On the other hand, range on the y-axes are chosen to cover the minimum and the maximum of the observed $\gamma$-ray fluxes since the launch of \fermi~satellite. Therefore, it is possible that we may not see the highest or the lowest $\gamma$-ray flux in the covered time period. The advantage of this approach is that it shows us the relative activity state of the source during our optical monitoring campaign. In Figure \ref{fig_gamma1}, we denote the epoch of the night observations with vertical lines. In order to understand the optical-\gm~ray connection for the covered period, we also utilize the optical $R$ band data taken from the Steward observatory \citep[][]{2009arXiv0912.3621S}, the Katzman Automatic Imaging Telescope \citep[KAIT;][]{2003PASP..115..844L}, and the SMARTS observatory \citep[][]{2012ApJ...756...13B}. Out of 17 sources, we could find the long term optical data for 10 sources. We correct the data for the galactic reddening following \citet[][]{2011ApJ...737..103S} and convert to flux units using the zero points of \citet[][]{1998A&A...333..231B}. The derived fluxes are shown in Figure \ref{fig_gamma1} with blue empty circles. It should be noted that adding the long term optical data was necessary since our INOV monitoring epochs are comparatively smaller for individual sources. We, then, perform a correlation test using the {\it z-transformed discrete correlation function} tool of \citet[][]{2013arXiv1302.1508A} to quantify the strength of the correlation in the optical and the \gm-ray bands. The derived results are presented in Table \ref{tab:dcf}. As can be seen that a strong claim cannot be made owing to the large errorbars and the small number of the data points and the results are consistent with the zero lead/lag. However, visual inspection of the light curves in Figure \ref{fig_gamma1} supports the idea that a high \gm-ray state is indeed accompanied by an optical flare and probably reflected in our INOV observations. Below, we individually discuss the results obtained for every blazar.

{\bf 3FGL J0608.0$-$0835:}\\
This source was observed twice, first on 2015 March 7 and then later on 2016 January 3 (Figure \ref{fig_inov1}). Both $C$ and $F$ tests have confirmed the absence of flux variations on the former epoch but a significant INOV of the order of $\sim$0.1 mag is noticed during the later observations. Along with this, an extremely fast ($<10$ minutes), small amplitude flare is also noticed at the beginning of the observation. To our knowledge, this is the first report of the detection of INOV from this object. The interesting fact to note is that during both the observing epochs, the source has not shown any significant $\gamma$-ray variability (see the top panel of Figure \ref{fig_gamma1}).

{\bf 3FGL J0656.4+4232:}\\
We have observed this object on 2016 January 5 but barring a couple of flickering events, no significant variability is detected.

{\bf 3FGL J0710.5+4732:}\\
This is the highest redshift source in our sample ($z=1.29$) and we have monitored it on three nights. Other than a few flickering, we have not found any significant INOV on any night, which is further confirmed from $C$ and $F$ statistics. During our monitoring campaign, the source remained quiescent in $\gamma$-rays , thus indicating the lack of the jet activity as a probable cause of the non-detection of INOV.

\begin{figure*}
\hbox{
\includegraphics[width=5.5cm]{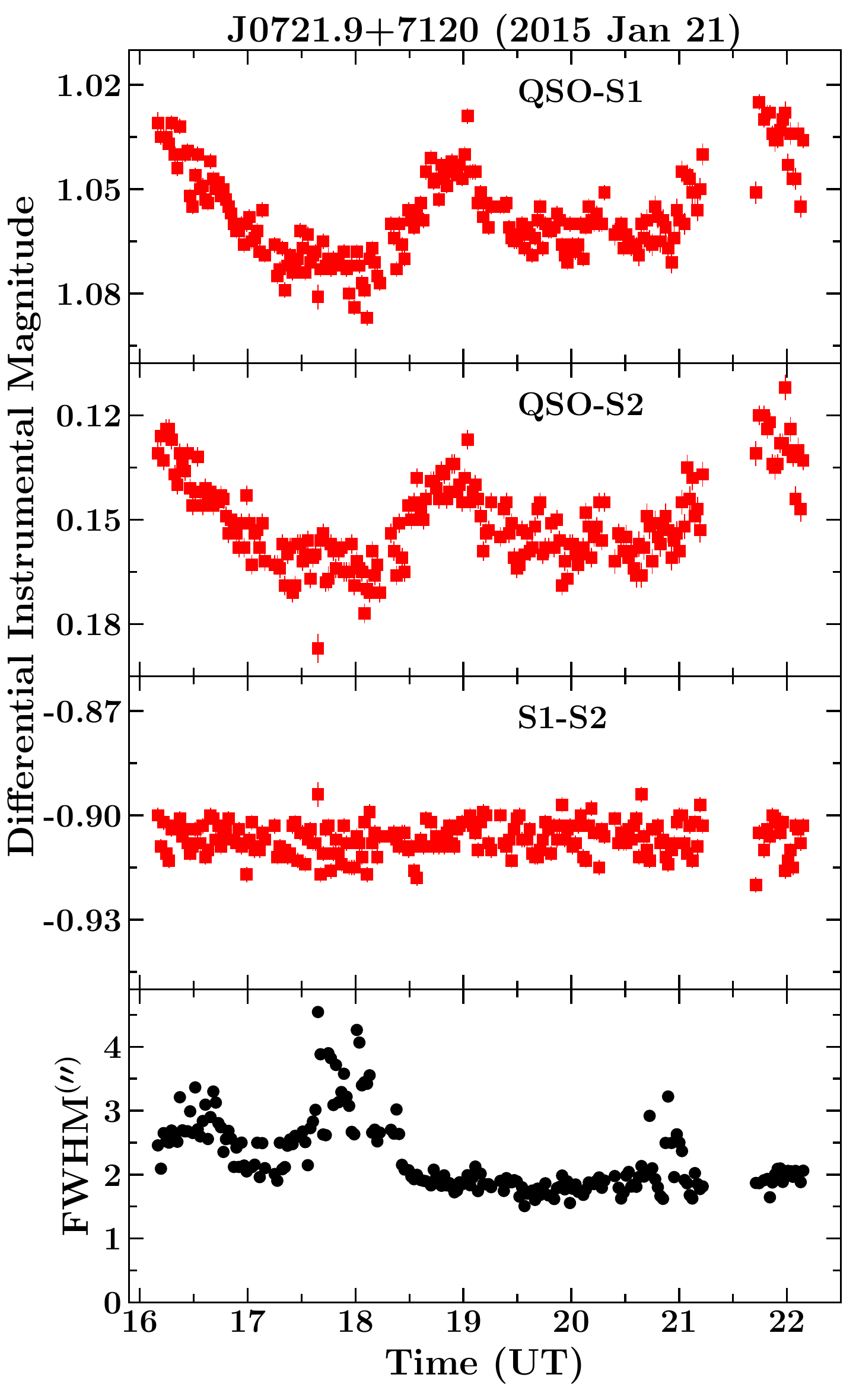}
\includegraphics[width=5.5cm]{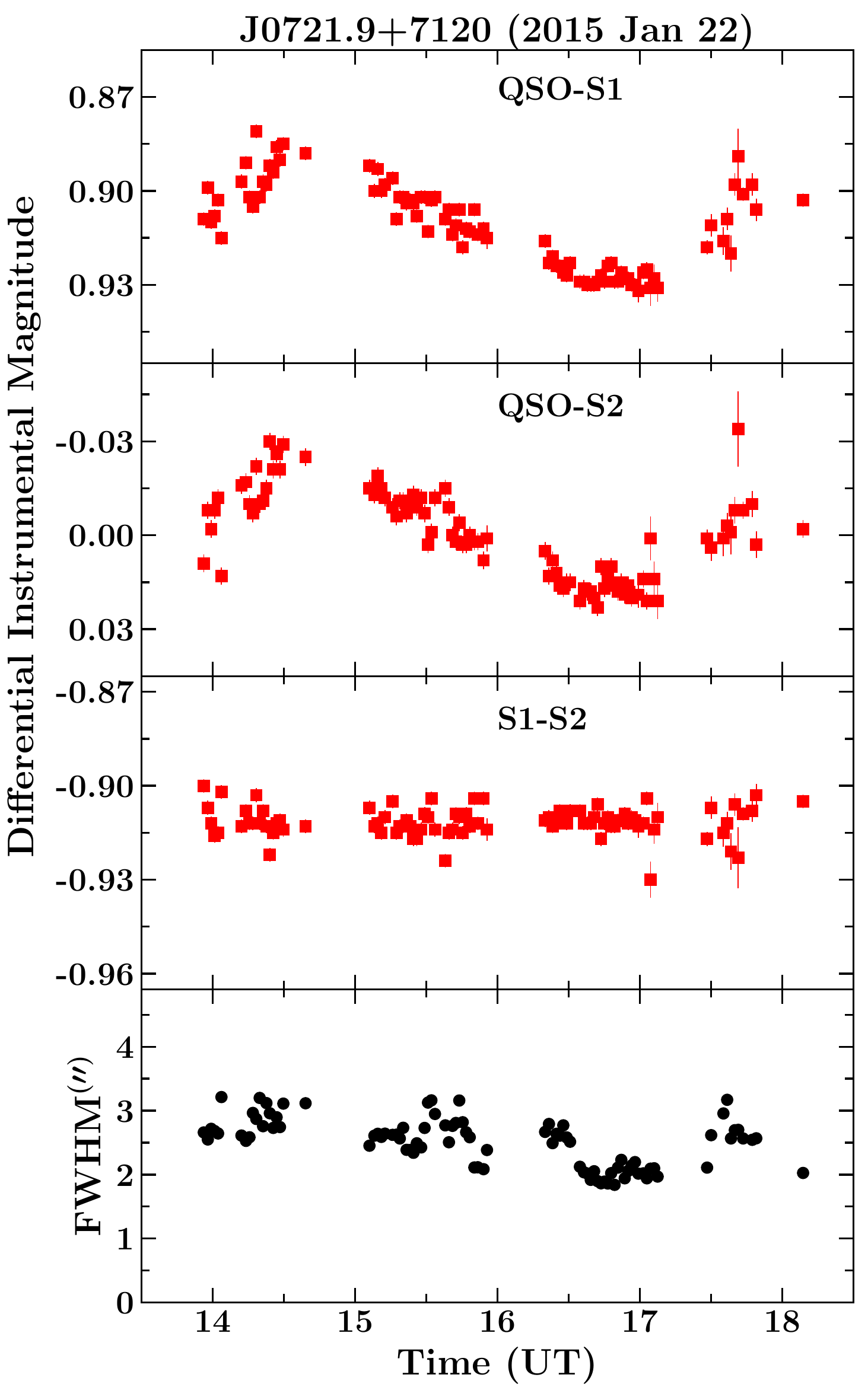}
\includegraphics[width=5.5cm]{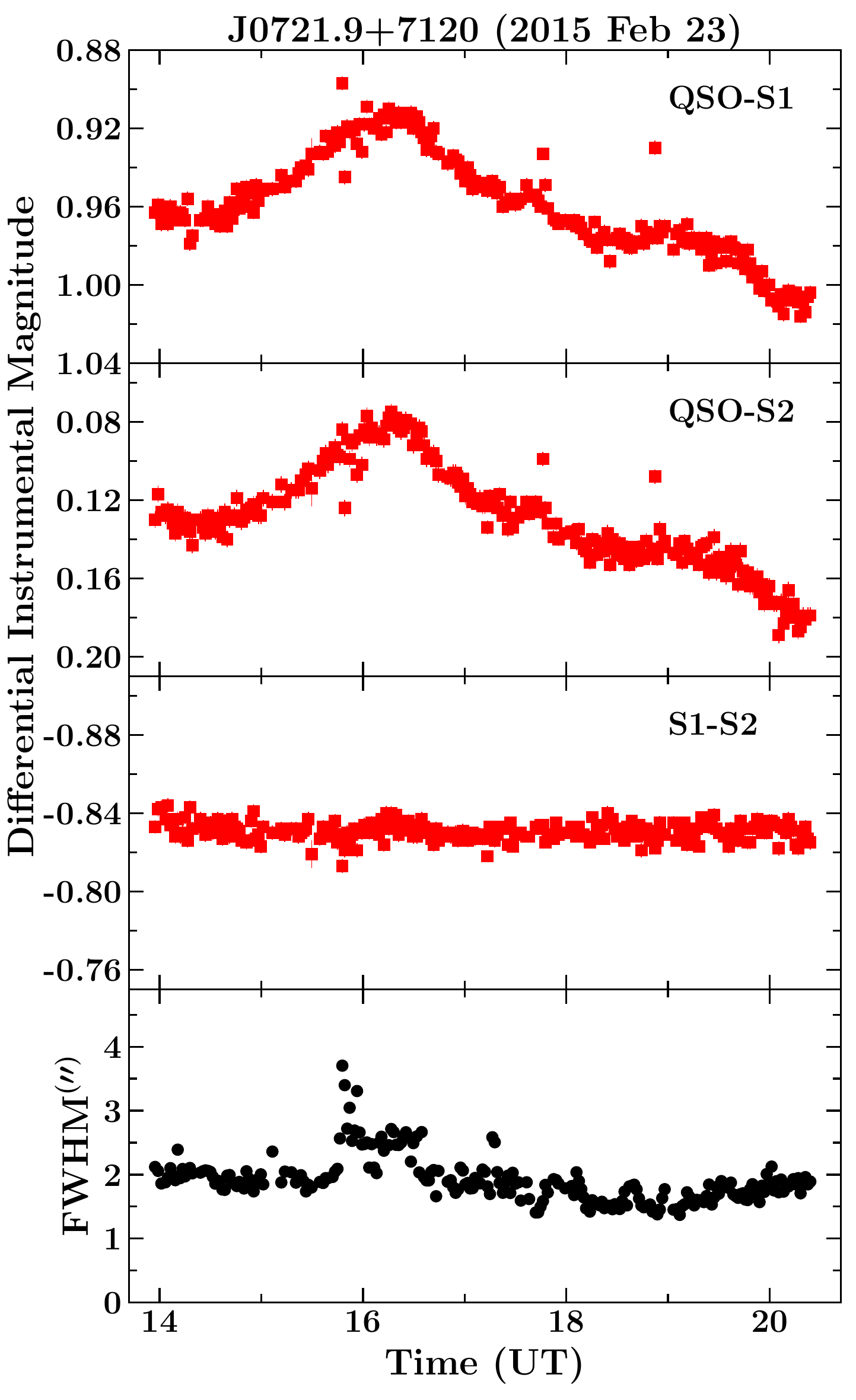}
}
\hbox{
\includegraphics[width=5.5cm]{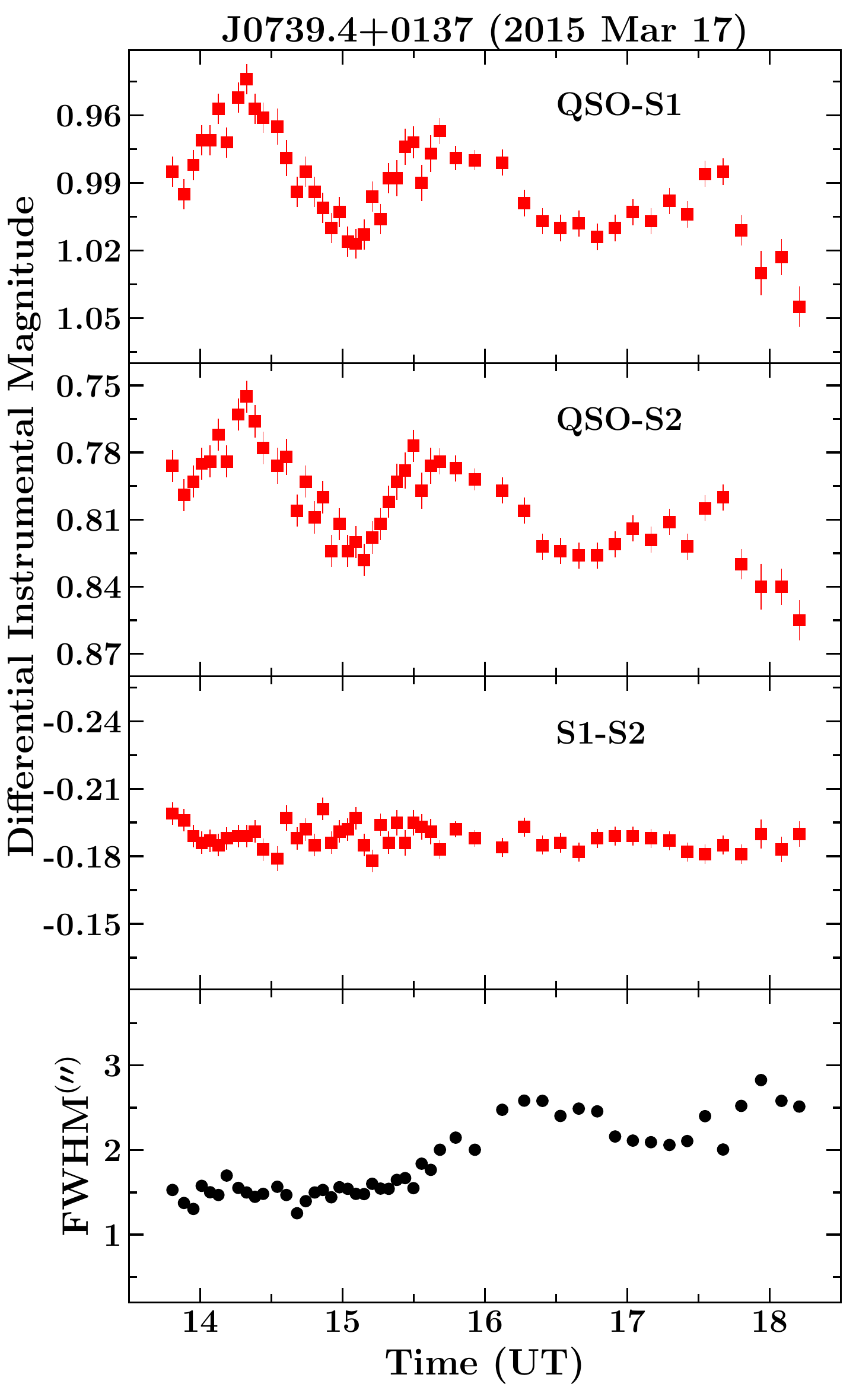}
\includegraphics[width=5.5cm]{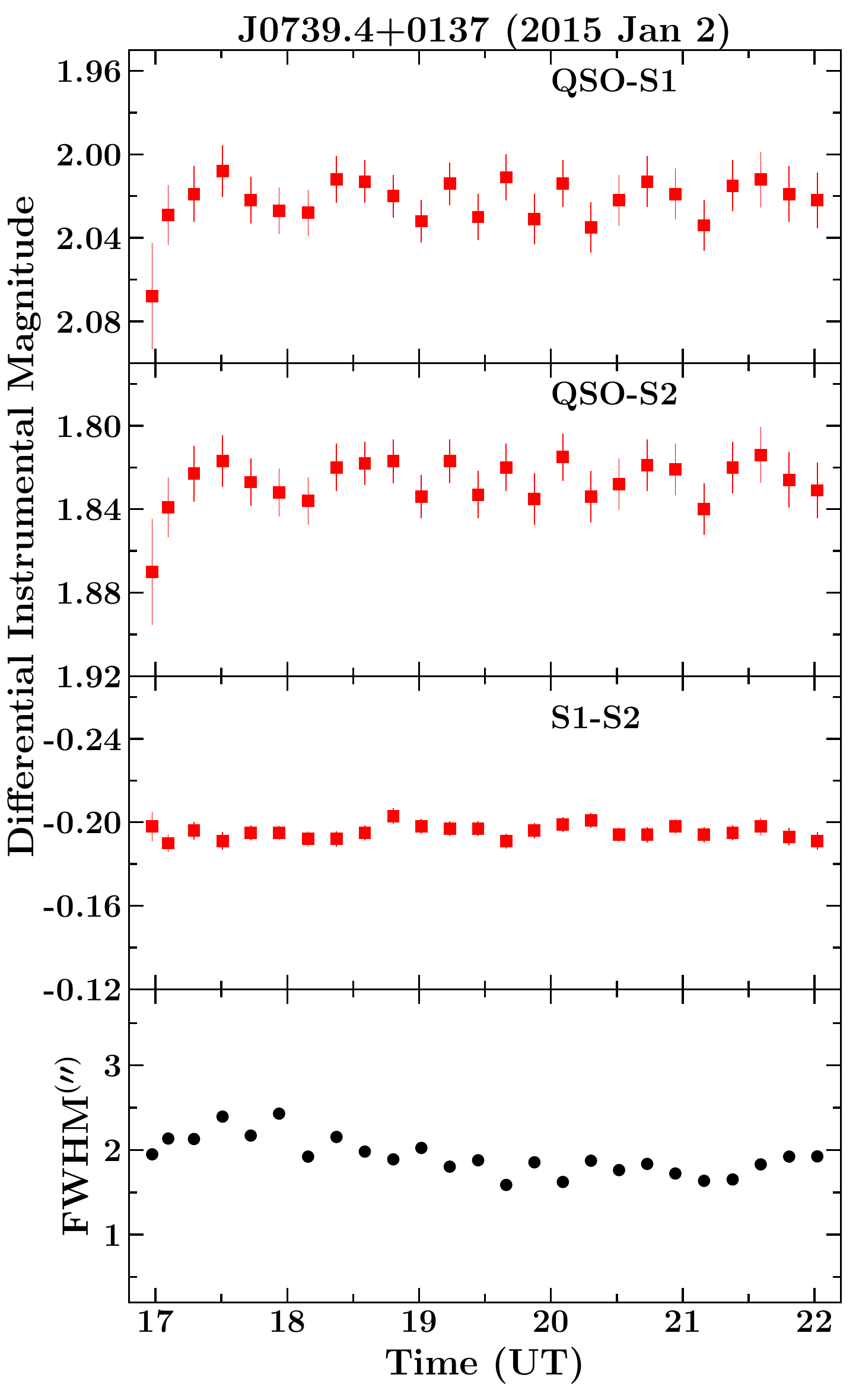}
\includegraphics[width=5.5cm]{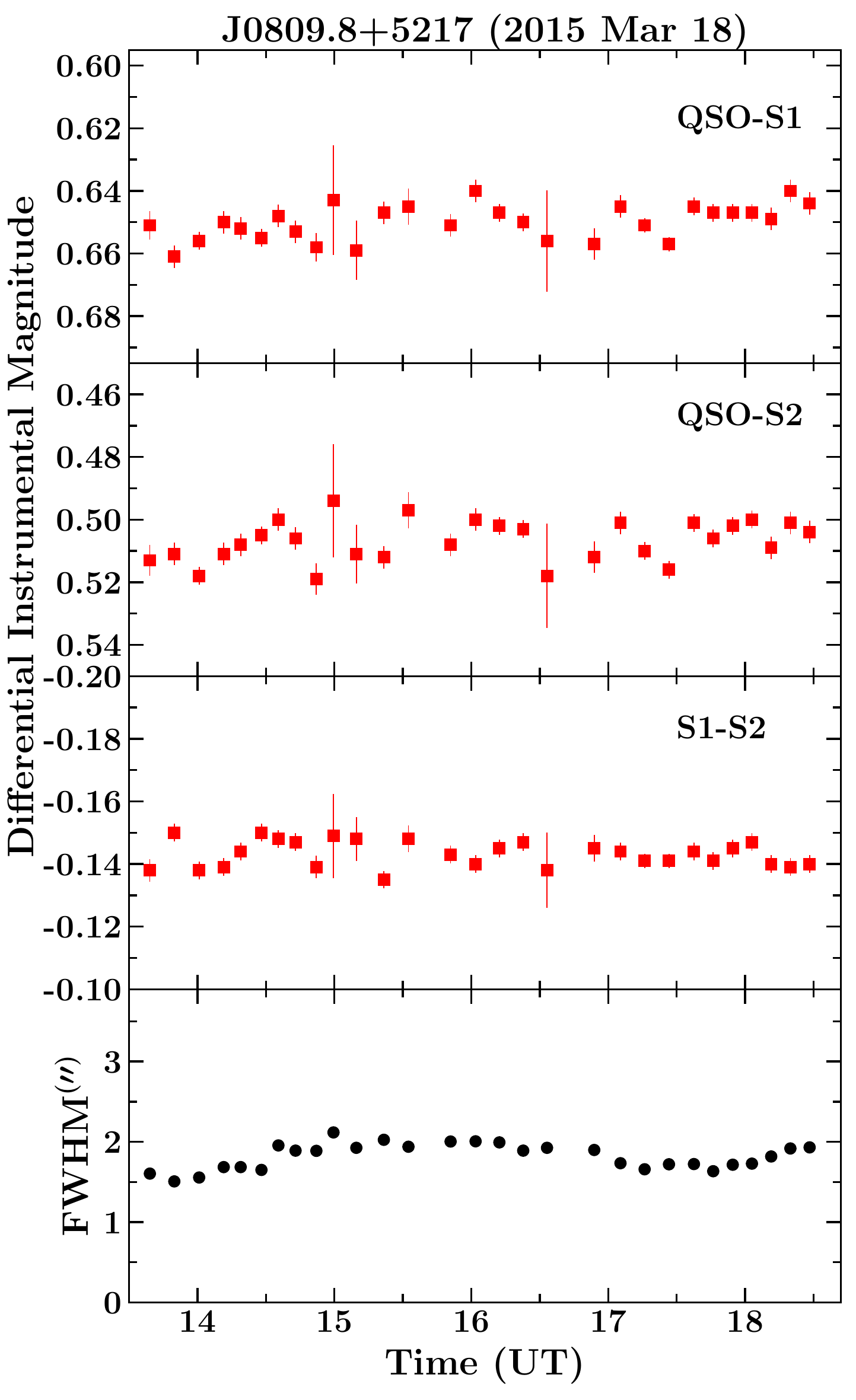}
}
\caption{Intra-night differential light curves of {\it Fermi} blazars. Other information are same as in Figure \ref{fig_inov1}.\label{fig_inov2}} 
\end{figure*}

{\bf 3FGL J0721.9+7120:}\\
This source is a well-known TeV blazar and it was observed on 2015 January 21, 22 and a month after on February 23. Our first two observations in 2015 January were coincident with the exceptional multi-wavelength flaring activity of this source \citep[Figure \ref{fig_gamma1}; see also, ][]{2015ApJ...809..130C,2015ATel.6999....1M,2015MNRAS.452L..11W,2016MNRAS.455..680A,2016MNRAS.458.2350W}. Such a high activity state is also reflected in its INOV behavior (Figure \ref{fig_inov2}). As can be seen in Figure \ref{fig_inov2}, a fast intra-night flare is noticed on 2015 January 21, that lasted for $\sim$1.5 hrs with the optical magnitude varies by about 0.04 mag. Similar behavior is also noticed during other two nights of the monitoring. Moreover, the source was also found to be in a high optical state as revealed from the Steward observatory monitoring. The fast optical variability from this source can be well explained from the beamed synchrotron emission which is further supported from the detection of the high optical polarization, $\sim5-10\%$, during the flaring episode from Steward observatory (see Table \ref{tab:stat}).

{\bf 3FGL J0739.4+0137:}\\
This is a FSRQ and is known to show fast INOV \citep[][]{2003AJ....126...37C}. \citet[][]{2012MNRAS.424.2625B} suggested that this blazar shows high amplitude variability when close to the optical maximum.  We find various small amplitude flares during its observation on 2015 March 7 (Figure \ref{fig_inov2}). Interestingly, during this period of the elevated activity, a high optical polarization of $\sim5\%$ was recorded at Steward observatory. In the conventional one-zone leptonic emission models, a high activity in the optical band is also reflected at high energy $\gamma$-rays. Therefore, we also looked into the monthly binned aperture photometric $\gamma$-ray light curve of this source to determine the association of the optical activity with $\gamma$-rays. As can be seen in Figure \ref{fig_gamma1}, the blazar was flaring in $\gamma$-rays in the month of 2015 March (see the first vertical line around MJD 57100), coinciding with the detection of the INOV. The similar results are also noticed in the long term optical data from the Steward observatory, though it appears that the source brightens even further after our INOV epoch. On the other hand, during the second night of the observation in 2016 January, when no significant optical variability was detected, it was in quiescence in the $\gamma$-ray band also. Therefore, it gives a strong indication that the probability of detecting a significant INOV from a blazar is high during the elevated $\gamma$-ray activity state. 

{\bf 3FGL J0809.8+5218:}\\
This source is a known TeV BL Lac object \citep[][]{2009ApJ...690L.126A} and its long term optical monitoring has been studied by \citet[][]{2014AJ....148..110M}. They did not find any significant INOV during the period spanned over $\sim$6 years. Our one night of the optical monitoring has not revealed any significant flux variability and we find this object to be in a very low $\gamma$-ray activity state during the epoch of the optical observation. The long term optical behavior of this object, as monitored by KAIT, also indicates the source to be in quiescence.

\begin{figure*}
\hbox{
\includegraphics[width=5.5cm]{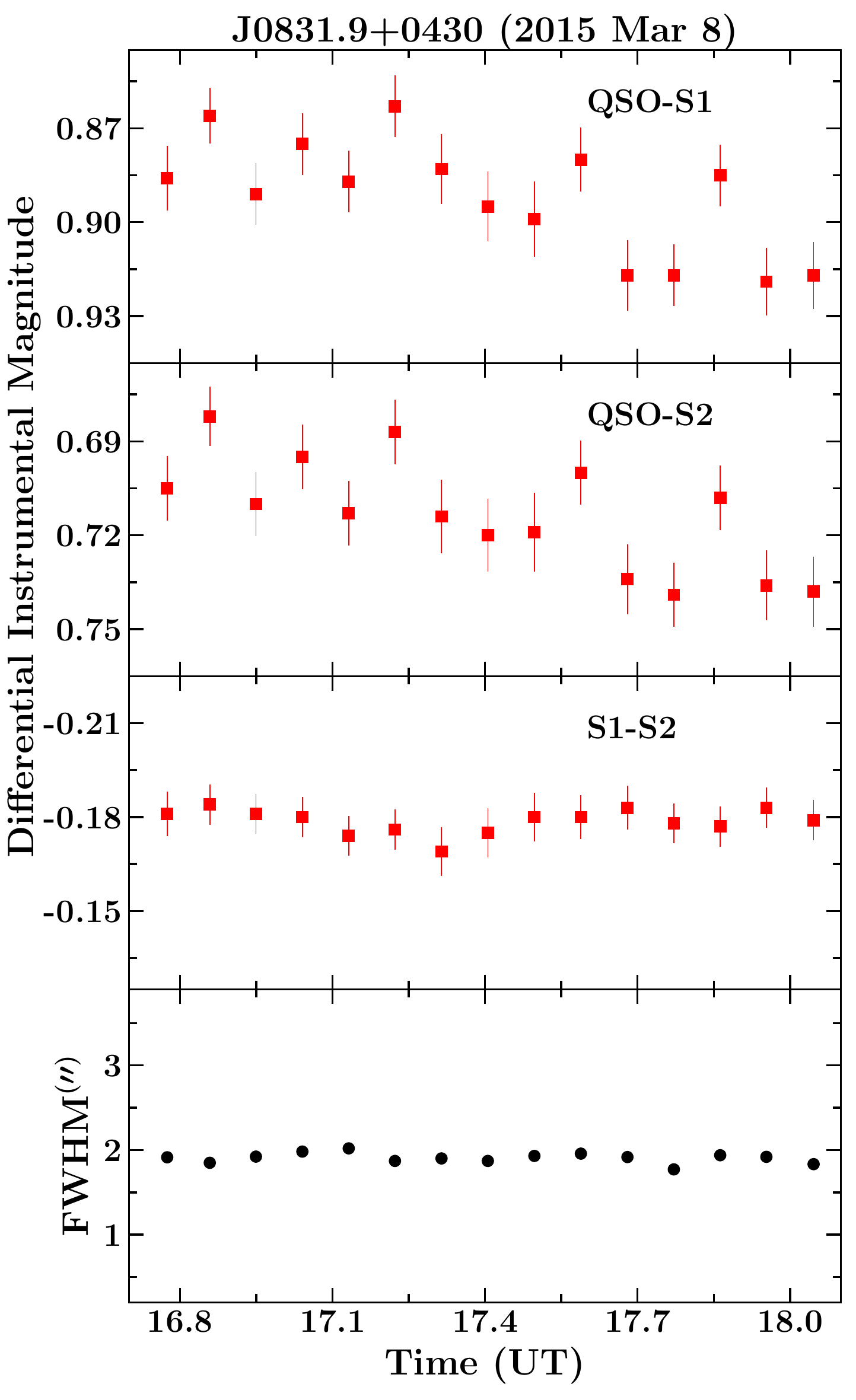}
\includegraphics[width=5.5cm]{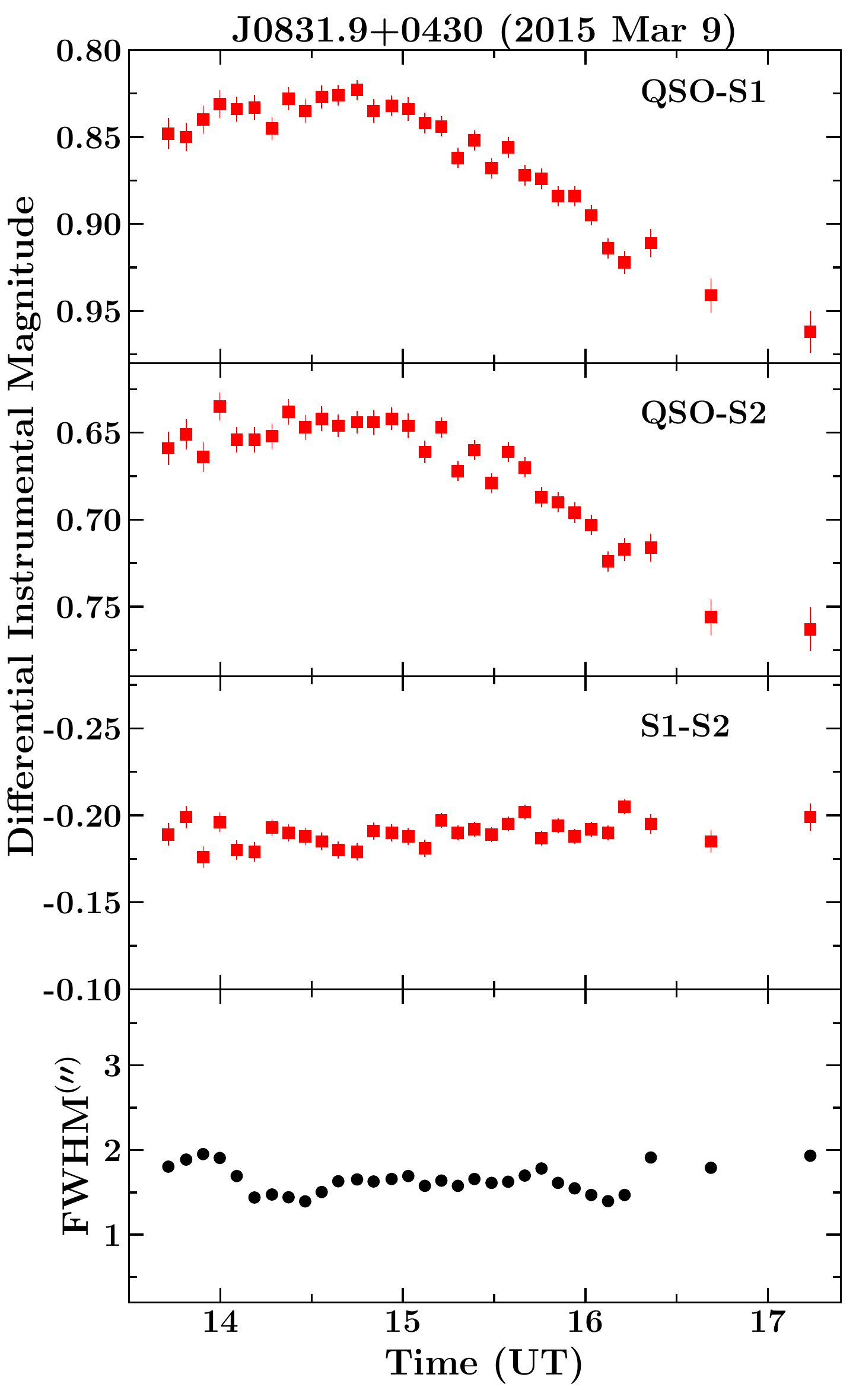}
\includegraphics[width=5.5cm]{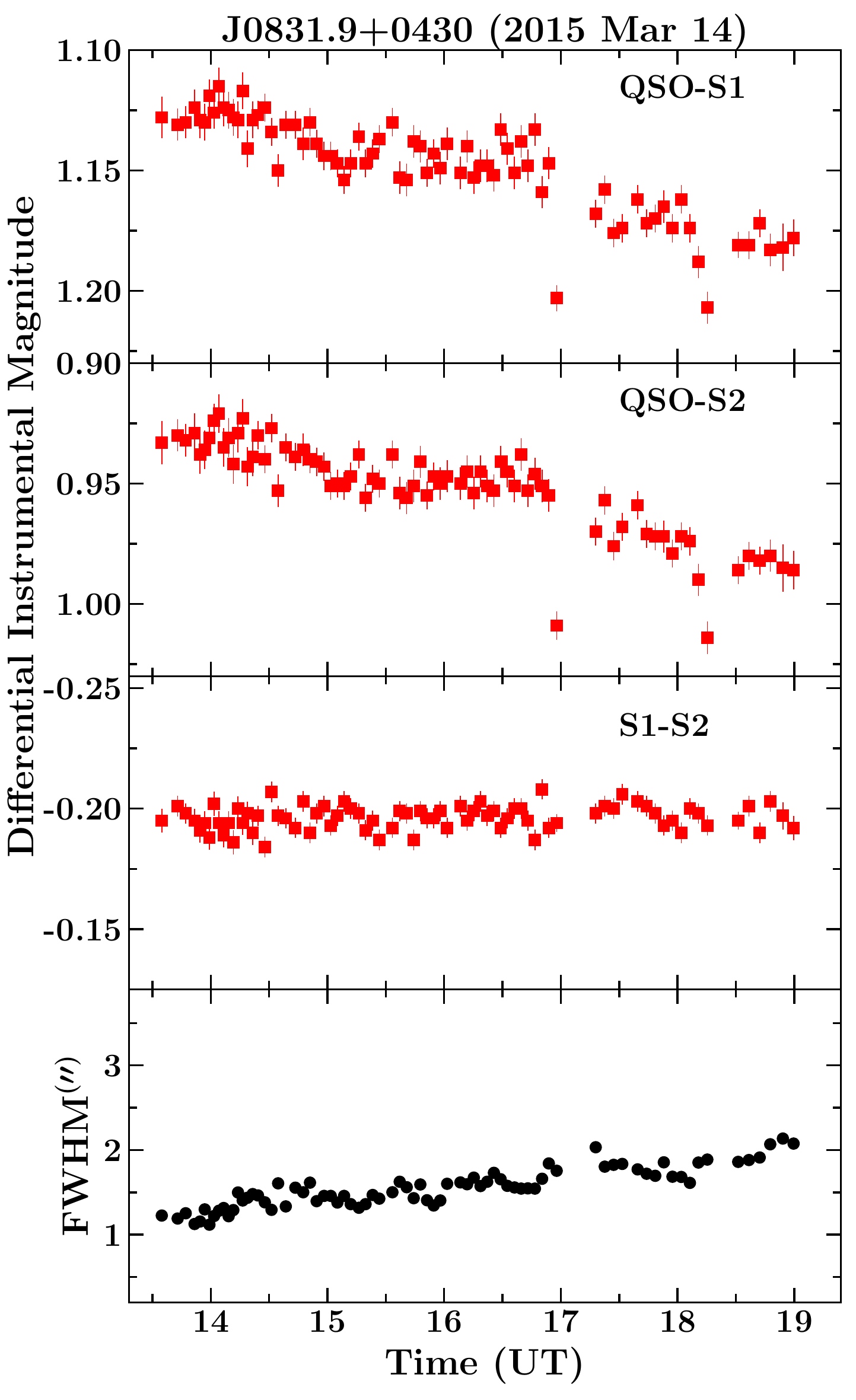}
}
\hbox{
\includegraphics[width=5.5cm]{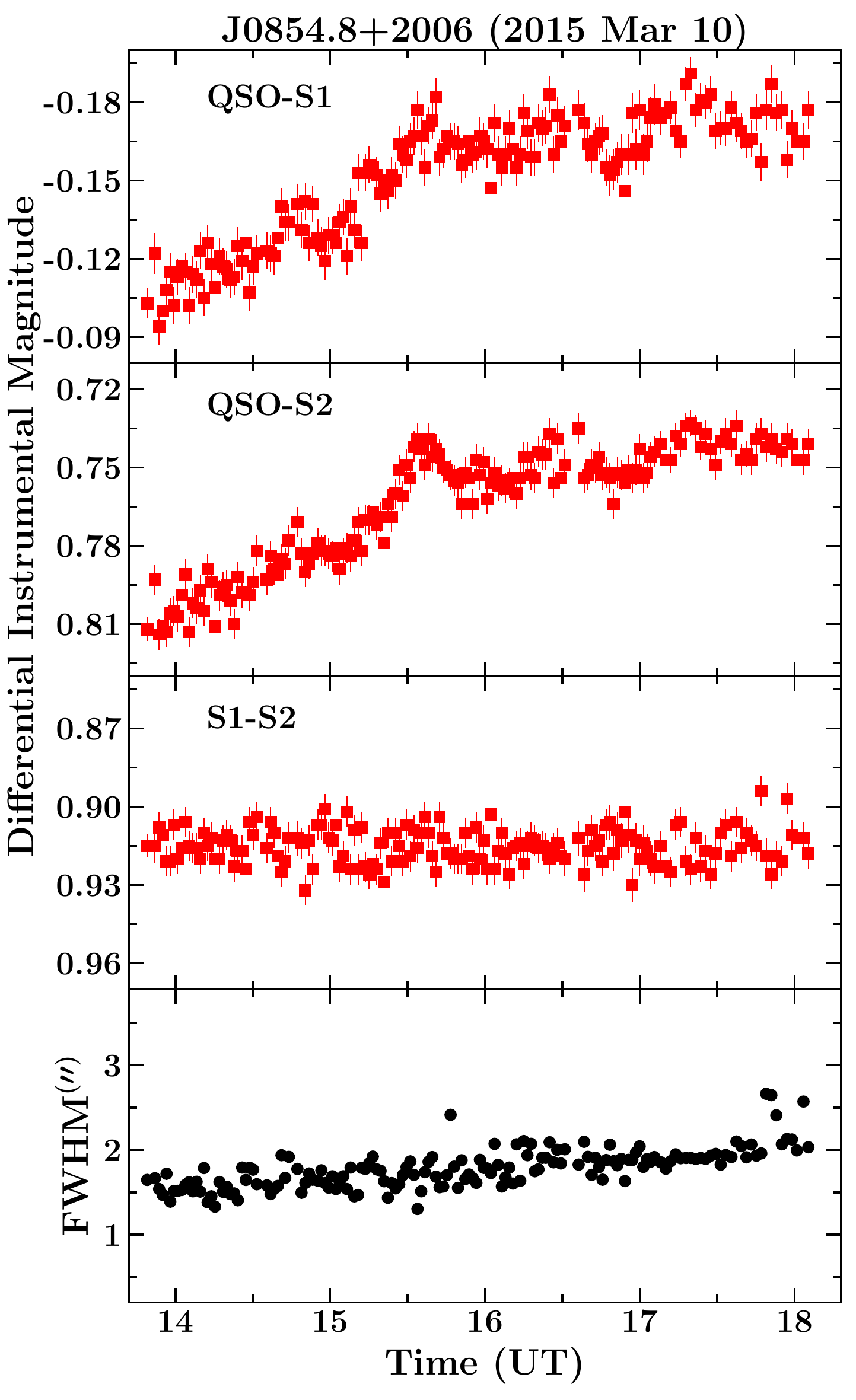}
\includegraphics[width=5.5cm]{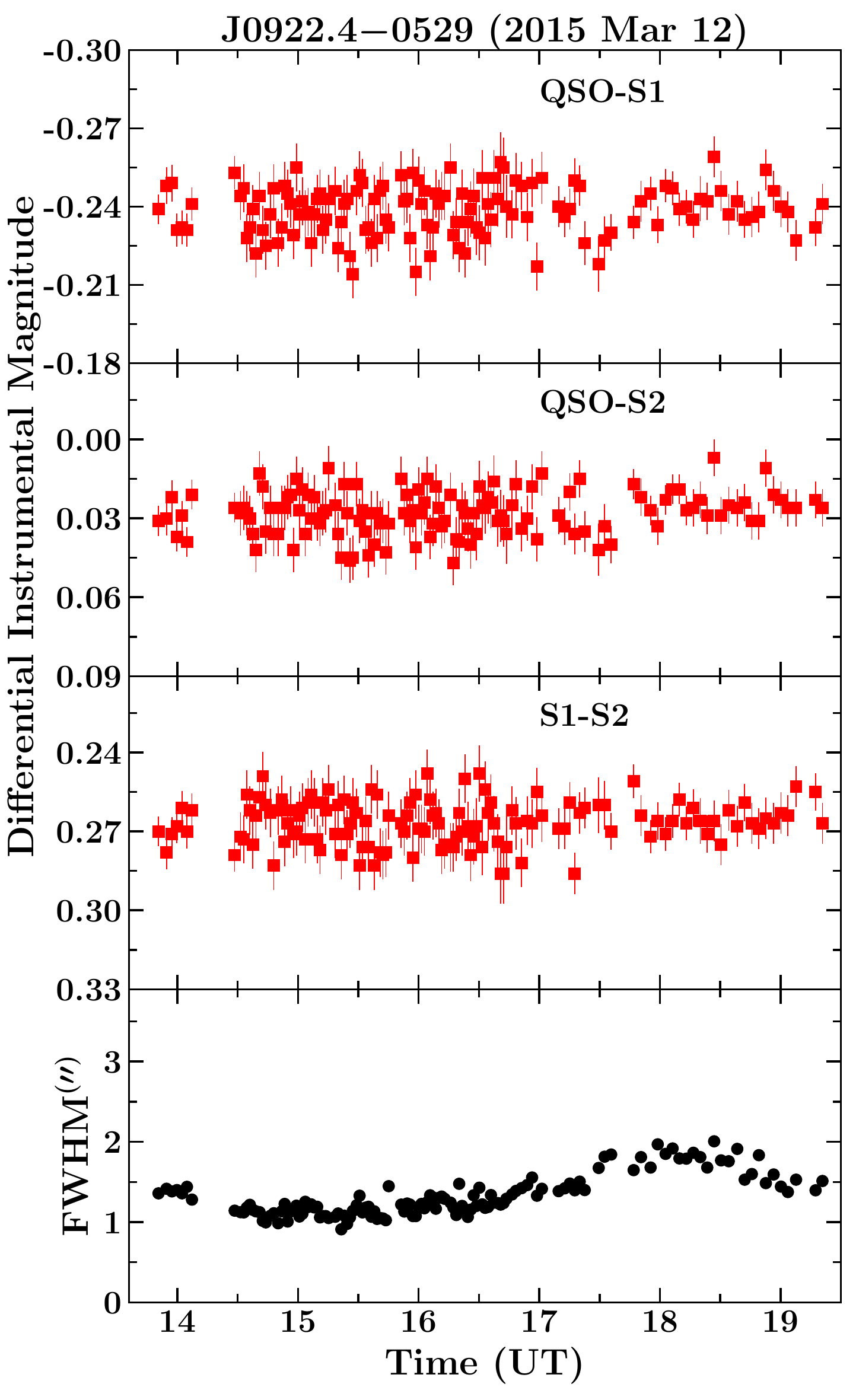}
\includegraphics[width=5.5cm]{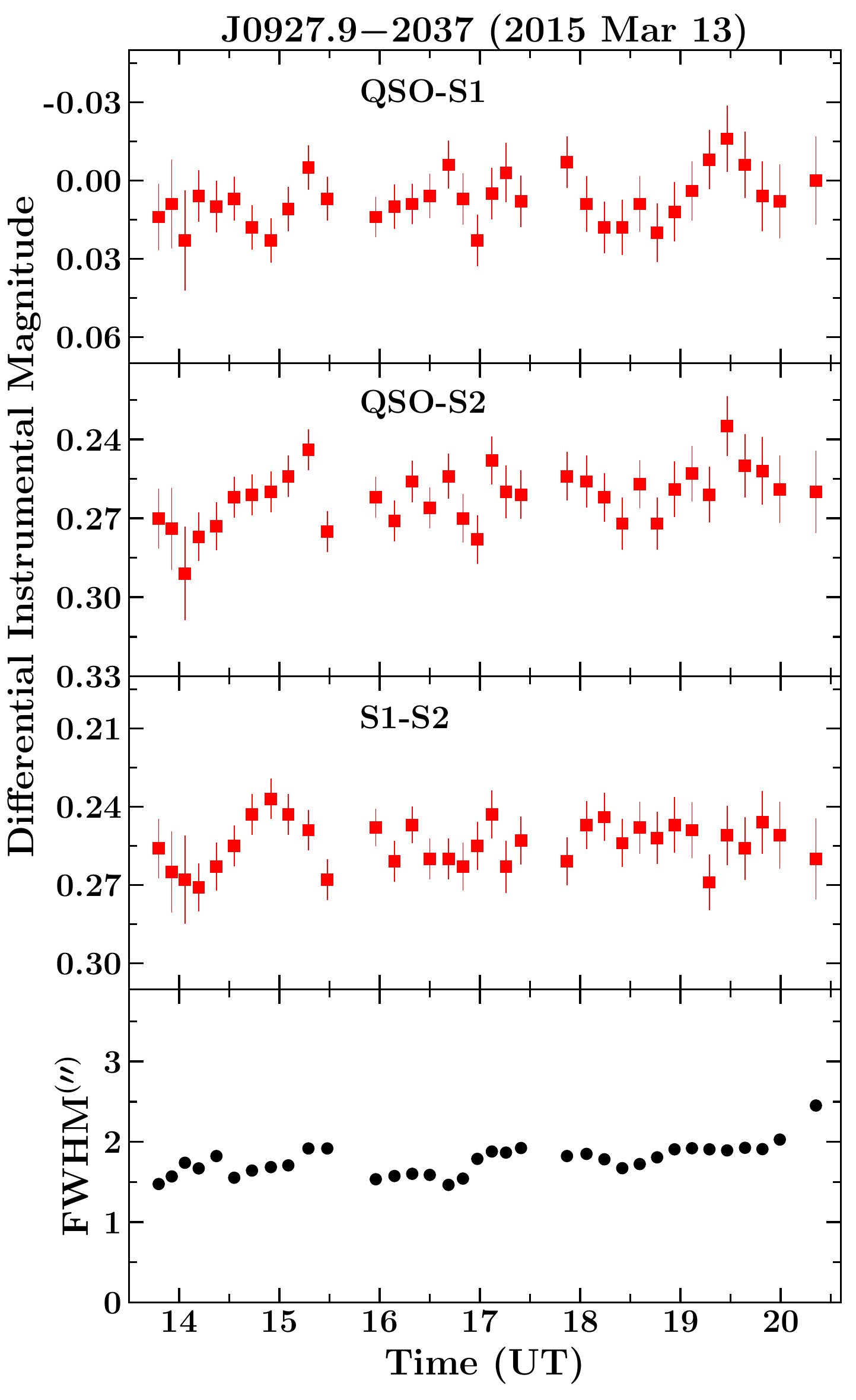}
}
\caption{Intra-night differential light curves of {\it Fermi} blazars. Other information are same as in Figure \ref{fig_inov1}.\label{fig_inov3}} 
\end{figure*}

{\bf 3FGL J0831.9+0430:}\\
3FGL J0831.9+0430 or PKS 0829+046 is a $\gamma$-ray luminous BL Lac object and it is observed at three different epochs. A significant INOV is detected during all three nights (Figure \ref{fig_inov3}) and this is probably the first report of the intra-night flux variations from this source. It is important to note that the source was $\gamma$-ray quiescent during our monitoring campaign. However, its IR-optical-UV spectrum is dominated by the non-thermal synchrotron emission \citep[][]{2011MNRAS.414.2674G} and hence the observed INOV could be originated from it. The long term KAIT monitoring reveals a moderate activity in the optical band during our observing run.

{\bf J0854.8+2006:}\\
OJ 287 is one of the best studied BL Lac objects in our sample and is known to show violent optical variability \citep[see, e.g.,][]{2009ApJS..181..466F,2011MNRAS.416..101G}. We have observed this source on 2015 March 10 and a significant INOV was detected. In particular, the source was found to be in a relatively moderate $\gamma$-ray activity state and we noted a small amplitude ($\sim$0.05 mag) flare that lasted less than an hour (see Figure \ref{fig_inov3}). The optical polarimetric monitoring from Steward observatory on 2015 March 17, closest to the epoch of our observation, reveals the detection of a high polarization of 13.4\% (Table \ref{tab:stat}), thus indicating the synchrotron based origin of the INOV.

{\bf 3FGL J0922.4$-$0529, 3FGL J0927.9$-$2037, 3FGL J1006.7$-$2159, 3FGL J1015.0+4925, and 3FGL J1204.3$-$0708:}\\
The blazars 3FGL J0922.4$-$0529, 3FGL J0927.9$-$2037, and 3FGL J1015.0+4925 were monitored for one night each and we observed 3FGL J1006.7$-$2159 and 3FGL J1204.3$-$0708 for two and three nights, respectively. None of them have shown any significant INOV (Figure \ref{fig_inov3}, \ref{fig_inov4}) and they were also found to be in $\gamma$-ray quiescence during the observing run (Figure \ref{fig_gamma1}).

{\bf 3FGL J1229.1+0202:}\\
We have monitored 3C 273 for three nights but could not detect any significant INOV (see Figure \ref{fig_inov5}). In fact, the source was in one of its lowest $\gamma$-ray and optical states during our observing campaign (see Figure \ref{fig_gamma1} and \url{http://quasar.square7.ch/fqm/1226+023.html}). It is also important to note that the optical spectrum of this object is dominated by a luminous accretion disk \citep[e.g.,][]{2015MNRAS.452.1303P} which is not expected to vary over short timescales ($\sim$few hrs). The detection of the almost negligible optical polarization ($<1$\%) recorded at Steward observatory and its historical micro-variability measurements \citep[][]{2001AJ....122.2901D} indicates that the lack of the INOV detection could be due to the dominance of the luminous accretion disk emission and the low activity of the jet.

 {\bf 3FGL J1404.8+0401:}\\
We have observed this object on 2015 May 20 for about 4.5 hrs and did not see any significant INOV.

{\bf 3FGL J1512.8$-$0906:}\\
We have monitored PKS 1510$-$08 during its exceptional flare in 2015 May (May 21 and 22). During this period of the high activity, it was significantly detected at very high energies (VHE, $E>100$ GeV) by MAGIC telescopes and the elevated activities were seen across the electromagnetic spectrum \citep[][]{2016arXiv161009416M,2016arXiv161102098Z}, including the optical monitoring from the Steward observatory. The source is found to exhibit a large amplitude variability ($\psi>15$\%) on the second night of the observation (see Table \ref{tab:stat}). On the other hand, the length of the observations ($<2$ hrs) during the first night was not enough to detect a significant INOV. The coincidence of the detection of the INOV with a high $\gamma$-ray activity state (Figure \ref{fig_gamma1}) again indicates a close association of detecting a high amplitude intra-night temporal variability during the high $\gamma$-ray state \citep[see also][for similar results]{2014ApJ...789..143P,2016ApJ...819..121P}. Similar to other blazars, where INOV was detected, the optical polarimetric monitoring from Steward observatory during the flaring period reports the detection of the high polarization. Interestingly, on 2015 May 21, i.e., when we could not detect INOV, the noticed optical polarization was 7.4\% and next day it increases to 15.3\%. This indicates that on 2015 May 22, PKS 1510$-$08 was in even higher optically flaring state which resulted in the detection of a significant micro-variability in our $\sim$2.5 hrs of the night observations.

{\bf 3FGL J1653.9+3945:}\\
Mkn 501 is a bright BL Lac object and it is known to display extremely fast TeV flux variations \citep[][]{2007ApJ...669..862A}. The source was in a low $\gamma$-ray and optical activity stated during our observing run. It is found to exhibit a significant flux variability on one of the monitoring nights. In fact, we detected an intriguing feature, a dip in the brightness of the source on $\sim$30 min timescale, on 2015 May 20 (Figure \ref{fig_inov6}). Such features have also been observed from another TeV blazar PG 1553+113 \citep[][]{2011MNRAS.416..101G} and are interpreted as possibly due to absorption in a dusty gas cloud, occulting the emission region responsible for the optical emission. The optical polarization of the source was quite low ($<3$\%) during both epochs.

\section{Discussion and Summary}{\label{sec:summary}}
We present the first results of our ongoing campaign to characterize the INOV properties of \fermi~detected blazars. This consists of the monitoring of 17 objects, 7 FSRQs and 10 BL Lac objects, for a total of $\sim$137 hrs. In our sample, BL Lac objects are found to display significantly higher DC of (40\%) than FSRQs ($\sim$15\%), considering $C$ and $F$ statistics. The conclusion that BL Lac objects show high INOV compared to FSRQs is also reported in various earlier studies \citep[e.g.,][]{2004MNRAS.348..176S}. Altogether, the overall DC is found to be $\sim$25\% which is well below than that known from blazars \citep[$\sim$70\%,][]{2004JApA...25....1S}. However, this is most probably due to the small sample size and less number of observations and it is likely that DC may further increase as we will acquire more observations. Alternatively, we assume a blazar to be variable when its amplitude of variability $\psi>3$\%, the overall DC increases to 66\% and it is 52\% and 86\% for FSRQs and BL Lac objects, respectively.  

\begin{figure*}
\centering
\hbox{
\includegraphics[width=9.0cm]{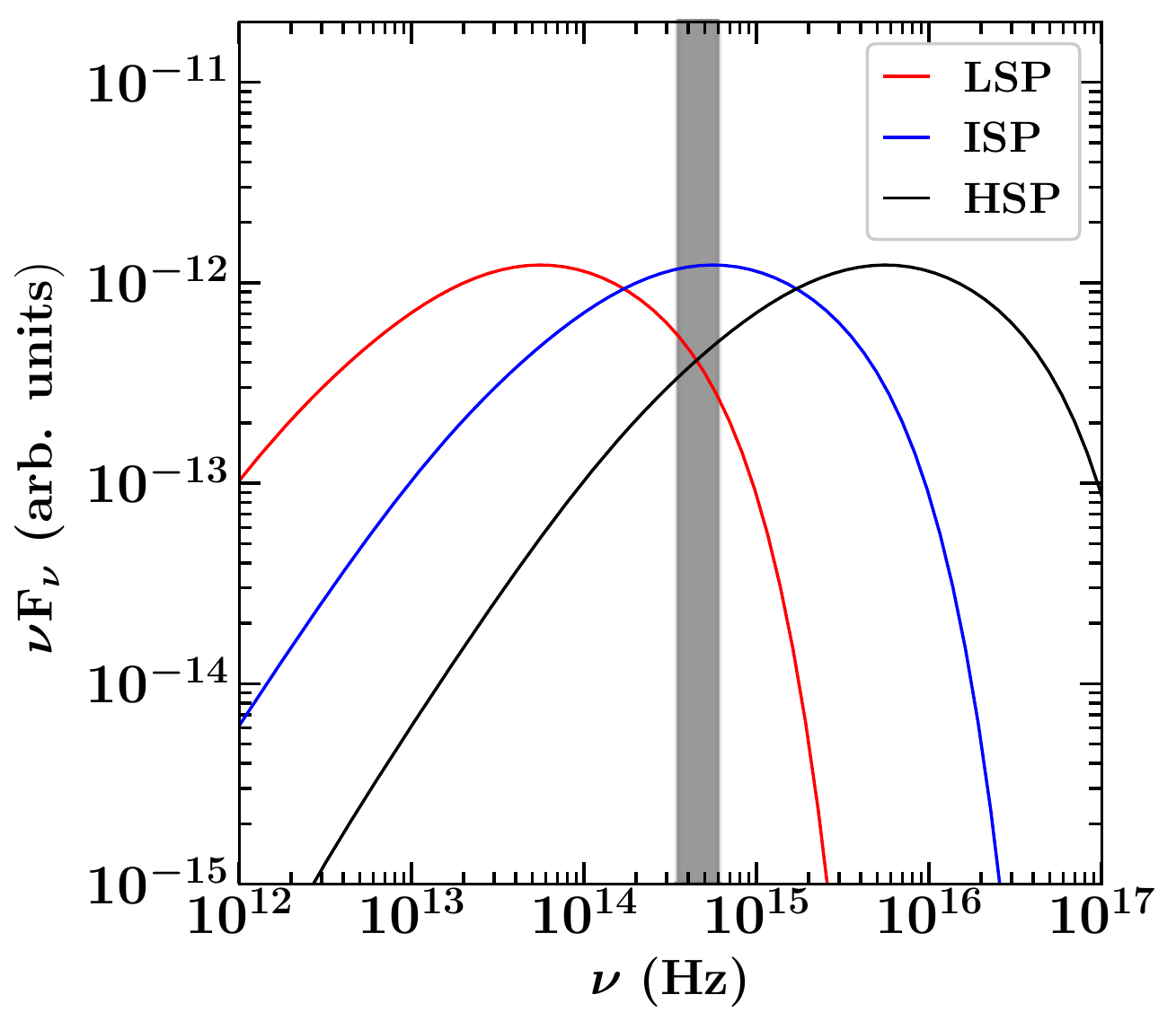}
\includegraphics[width=8.2cm]{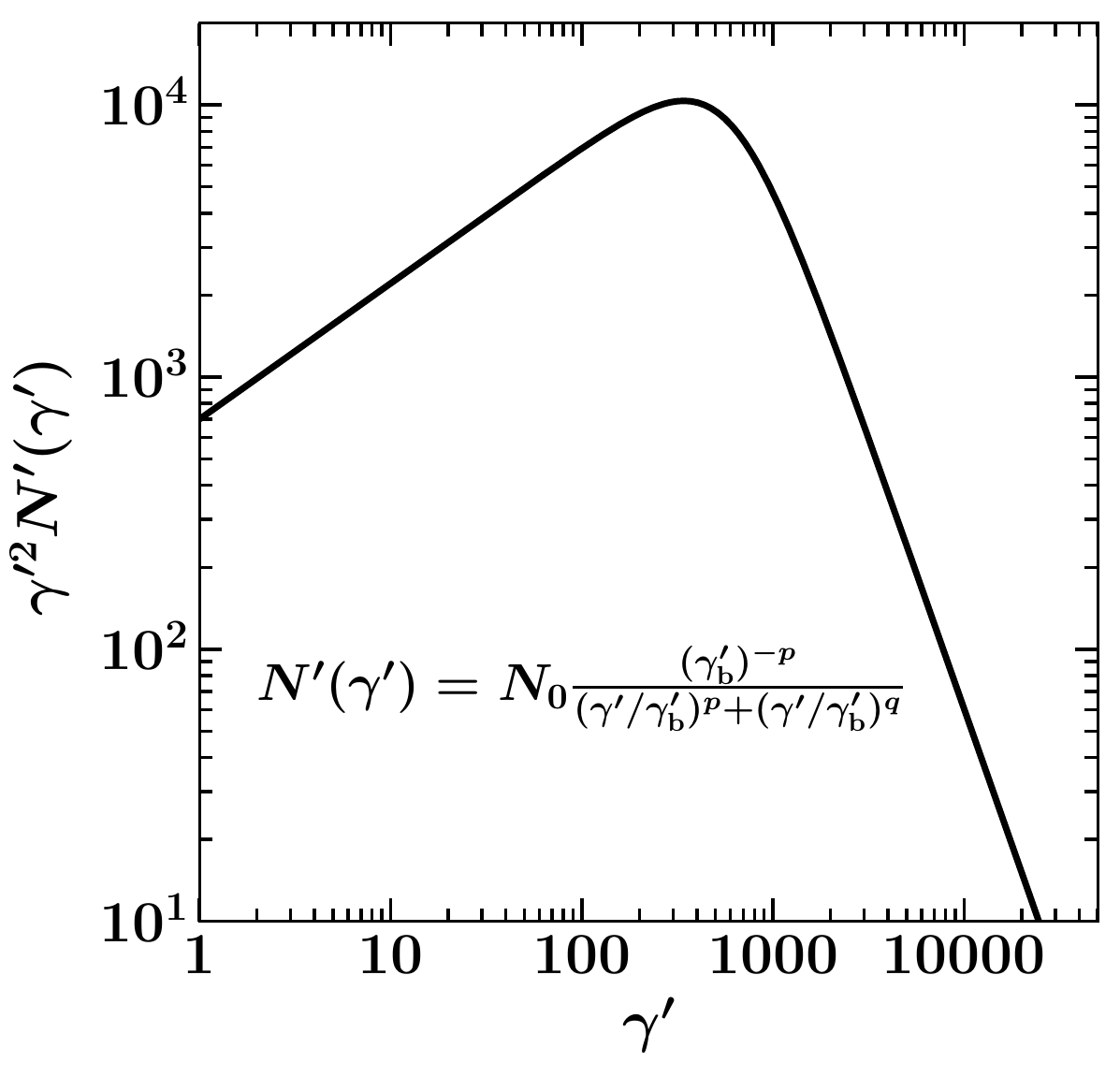}
}
\caption{ Left: A typical synchrotron spectrum of LSP (red), ISP (blue), and HSP (black) blazars. The grey strip represents the wavelength range covered in the $R$ band. Right: The energy distribution of the relativistic electrons in the conventional leptonic emission scenario, as described by a smooth broken power law model. The parameter $\gamma$ is the random Lorentz factor of the electrons, whereas, $p$ and $q$ denote the slopes of the model before and after the peak energy ($\gamma_{\rm b}$). The primed quantities are in the comoving frame of the emission region. The spectral parameters are appropriately chosen to show the particle energy distribution. Comparing with the left panel, it can be noticed that the rising part of the synchrotron spectrum is emitted by the low energy electrons, whereas, high energy electrons contribute to the falling part of the SED. \label{fig_syn}} 
\end{figure*}

On classifying the sources according to the location of the synchrotron peaks, we find that LSP blazars exhibit significantly larger variability compared to ISP and HSP objects. Though a strong claim cannot be made due to the small size of the sample, the following discussion can provide a possible physical explanation. In the left panel of Figure \ref{fig_syn}, we show a typical synchrotron spectrum \citep[e.g.,][]{2008ApJ...686..181F} of LSP (red), ISP (blue), and HSP (black) blazars and also plot the frequency range covered by the Bessel $R$ filter (grey strip), using which we have carried out the observations. As can be seen, for the case of LSP sources, $R$ band covers the falling part of the spectrum, whereas, ISP objects have their peak located at these energies. For HSP blazars, on the other hand, a rising synchrotron spectrum is covered by the $R$ band. In the conventional leptonic emission scenario, the broadband emission from blazars originates from the relativistic electron population whose energy distribution follows a broken power law spectral shape (see the right panel of Figure \ref{fig_syn}). This means that a rising synchrotron spectrum is produced by the low energy electrons, whereas, the high energy particles contribute to the falling part of the SED. In other words, for ISP and HSP blazars, a relatively low energy electron population radiates in the $R$ band compared to LSP sources. Since the cooling time is inversely proportional to the energy of the radiating particles ($t_{\rm cool}\propto \gamma^{-1}$, $\gamma$ is the Lorentz factor of the electrons), it is obvious that the high energy electrons should exhibit faster variability, implying LSP sources to be more variable compared to ISP and HSP blazars. In fact, HSP blazars display violent flux variability at hard X-rays and at TeV energies \citep[see, e.g.,][]{1996Natur.383..319G,2015ApJ...811..143P} where the highest energy electrons, forming the tail of the synchrotron and inverse Compton processes, radiates.

Our observing results provide evidences about a close association of the detection of the high amplitude INOV and a high optical polarization (monitored from Steward observatory) with the elevated $\gamma$-ray activity in blazars. This probably indicates a high chance of detecting large amplitude INOV when a blazar goes into the $\gamma$-ray flaring state. This feature has already been observed in a few $\gamma$-ray emitting narrow line Seyfert 1 galaxies \citep[][]{2014ApJ...789..143P,2016ApJ...819..121P} and therefore, it appears reasonable to generalize it to the beamed $\gamma$-ray emitting AGNs. This also supports the jet based origin of the flares rather than from the accretion disk. Furthermore, we also find extremely fast flux variations of the orders of minutes (e.g., see the light curves of 3FGL J0608.0$-$0835). Such rapid flares are typically observed from blazars at VHE $\gamma$-rays and they are probably originated from the magnetic reconnection events \citep[][]{2009MNRAS.395L..29G}, structured jets \citep[][]{2008AIPC.1085..431T}, due to turbulence in the plasma flow \citep[e.g.,][]{2012MNRAS.420..604N,2014ApJ...780...87M}, or possibly due to rapid conversion of the magnetic energy to radiation by a process called {\it magnetoluminescence} \citep[][]{2015AAS...22521407B}. It is, therefore, of utmost interest to detect extremely rapid optical flares within the night to further test these theories of rapid flux variability detected from blazars. 

The easily accessible small 1-2 m class optical telescopes definitely play an important role in studying the fundamental problems of blazar physics. In future, apart from 1.3 m JCBT, we also plan to extend the observations of $\gamma$-ray blazars from various other small telescopes, such as 0.9 m SARA-North at Kitt Peak, Arizona and 0.6 m SARA-South at Cerro Tololo, Chille, to verify/confirm the findings reported in this article and those existed in literature.

\acknowledgments
We are grateful to the referee for providing a constructive report. VSP and CSS are thankful to V. Moorthy, M. Appakutty, G. Selvakumar, S. Venkatesh, B. Rahul, and P. Anbazhagan,  and the staff members of the VBO, for their assistance during the observing run. VSP acknowledge the warm hospitality received at VBO where this work was completed. This research has made use of data obtained from the Leicester Database and Archive Service at the Department of Physics and Astronomy, Leicester University, UK. Data from the Steward Observatory spectropolarimetric monitoring project were used. This program is supported by Fermi Guest Investigator grants NNX08AW56G, NNX09AU10G, NNX12AO93G, and NNX15AU81G. This paper has made use of up-to-date SMARTS optical/near-infrared light curves that are available at www.astro.yale.edu/smarts/glast/home.php.

\bibliographystyle{aasjournal}
\bibliography{Master}

\begin{table*}
\caption{The list of the $\gamma$-ray blazars  monitored in this work. Column information are as follows: (1) 3FGL name; (2) other name; (3) right ascension (J2000); (4) declination (J2000); (5) redshift;  (6) apparent $R$ band magnitude; (7) blazar type. All the information, except $R$ band magnitudes, are from \citet{2015ApJ...810...14A}. $R$ band magnitudes are adopted from USNO-B1 catalog \citep{2003AJ....125..984M}.\label{tab:basic_info}
}
\begin{center}
\begin{tabular}{llccccl}
\hline
3FGL name  & Other name & RA (2000) & Dec (2000) & $z$ & $R$ &  Type \\
(1) & (2) & (3)  & (4) & (5) &  (6)  & (7)\\ 
\hline															     			
J0608.0$-$0835 & PKS 0605$-$08 & 06 07 59.62  &  $-$08 34 52.16     &  0.872   &       15.41   & ISP FSRQ\\
J0656.4+4232    & 4C +42.22 & 06 56 10.67  & 42 37 02.90        &  0.059   &       11.11   & HSP BL Lac\\
J0710.5+4732    & S4 0707+47 & 07 10 46.31  & 47 32 13.46        &  1.292   &       13.44   & ISP BL Lac\\
J0721.9+7120    & S5 0716+71 & 07 21 53.39  & 71 20 36.65        &  0.127   &       14.14   & LSP BL Lac\\
J0739.4+0137    & PKS 0736+01 & 07 39 18.03  & 01 37 04.61        &  0.189   &       16.03   & ISP FSRQ\\
J0809.8+5218    & 1ES 0806+524 & 08 09 49.20   & 52 18 58.35       &  0.138   &       14.63   & HSP BL Lac\\
J0831.9+0430    & PKS 0829+046 & 08 31 48.88   & 04 29 38.73       &  0.174   &       13.99   & LSP BL Lac\\
J0854.8+2006    & OJ 287 & 08 54 48.87   & 20 06 30.83       &  0.306   &       14.35   & LSP BL Lac\\
J0922.4$-$0529 & TXS 0919$-$052 & 09 22 24.09   & $-$5 29 08.14      &  0.974   &       14.75   & ISP FSRQ\\
J0927.9$-$2037 & PKS 0925$-$203 & 09 27 51.83   & $-$20 34 51.47     &  0.348   &       15.98   & LSP FSRQ\\
J1006.7$-$2159 & PKS 1004$-$217 & 10 06 46.41   & $-$21 59 20.38     &  0.330   &       15.23   & ISP FSRQ\\
J1015.0+4925    & 1H 1013+498 & 10 15 4.13     & 49 26 00.87      &  0.212   &       15.13   & HSP BL Lac\\
J1204.3$-$0708 & 1RXS J120417.0$-$070959 & 12 04 16.66   & $-$7 10 09.16      &  0.184   &       14.59   & HSP BL Lac\\
J1229.1+0202    & 3C 273 & 12 29 06.70   & 02 03 08.59       &  0.158   &       14.09   & LSP FSRQ\\
J1404.8+0401    & MS 1402.3+0416 & 14 04 50.92   & 04 02 02.38       &  0.344   &       17.08   & HSP BL Lac\\
J1512.8$-$0906 & PKS 1510$-$08 & 15 12 50.54   & $-$09 05 59.69     &  0.360   &       15.70   & LSP FSRQ\\
J1653.9+3945    & Mkn 501 & 16 53 52.21   & 39 45 36.31       &  0.034   &        9.07   & HSP BL Lac\\
\hline
\end{tabular}
\end{center}
\end{table*}

\begin{table*}
\caption{Deatils of the CCDs mounted on the JCBT. The bottom two panels report the read out noise and gain of the proEM CCD at different operating frequencies.\label{tab:ccd}}
\begin{center}
\begin{tabular}{lcc}
\hline
Characteristics & proEM CCD  & normal CCD \\
\hline															     			
Array (pixels)                                                   & 1024 $\times$ 1024 & 2048 $\times$ 4096\\
Pixel size (micron)                                            & 13                             & 15                            \\
Plate scale	(arcsec pixel$^{-1}$)                     & 0.26                          & 0.30	\\
Full well capacity (e$^{-}$)                               & 8 $\times$ 10$^4$     &  2 $\times$ 10$^5$\\
Operating Temperature                                     & $-$68$^{\circ}$ C     &   $-$100$^{\circ}$ C\\
Dark current (e$^{-}$ pixel$^{-1}$ sec$^{-1}$) & 0.04                         &  $4\times10^{-4}$ \\
Read out noise (e$^{-}$)                                   & variable                     & 3.8 \\
Gain (e$^{-}$ ADU$^{-}$)                                 & variable                     & 0.74  \\
\hline
      & proEM CCD Gain & \\
Frequency	  & High gain          & Medium gain  \\
\hline
100 kHz        & 0.699               & 1.38            \\
1 MHz           & 0.693               & 1.37            \\
5 MHz           & 1.19                 & 2.26            \\
\hline
      & proEM CCD read out noise & \\
Frequency	  & High gain          & Medium gain  \\
\hline
100 kHz        & 3.87               & 4.74            \\
1 MHz           & 6.24               & 10.35            \\
5 MHz           & 13.0               &  14.20           \\
\hline
\end{tabular}
\end{center}
\end{table*}

\begin{table*}
\caption{Positions and $R$ band apparent magnitudes of the comparison stars, associated with the target blazars. The information are taken from the USNO catalogue \citep{2003AJ....125..984M}.\label{tab:com_star}}
\begin{center}
\begin{tabular}{lcccc}
\hline
Source & Star & RA (J2000) & Dec (J2000) & $R$ \\
 &  &  &  & (mag)\\
\hline
J0608.0$-$0835   & S1 & 06  07  54.01 & $-$08  33  37.23 &  15.22\\
                 & S2 & 06  08  00.59 & $-$08  35  51.63 & 15.18\\
J0656.4+4232     & S1 & 06  56  05.95 & 42   38  52.16   &  14.71\\
                 & S2 & 06  56  17.26 & 42  35   52.35   & 14.63\\
J0710.5+4732     & S1 & 07  10  53.52 & 47   30  13.35   &  12.89\\
                 & S2 & 07  10  38.60 & 47  33   14.51   & 12.80\\
J0721.9+7120     & S1 & 07  21  52.29 & 71   18  17.92   &  13.11\\
                 & S2 & 07  22  17.99 & 71  23   34.83   & 13.55\\
J0739.4+0137     & S1 & 07  39  16.14 & 01   37  34.93   &  15.02\\
                 & S2 & 07  39  23.38 & 01  35   28.24   & 15.15\\
J0809.8+5218     & S1 & 08  10  03.24 & 52   18  57.82   &  14.56\\
                 & S2 & 08  10  17.08 & 52  19   43.77   & 15.17\\
J0831.9+0430     & S1 & 08  32  04.43 & 04   25  52.38   &  14.44\\
                 & S2 & 08  31  47.41 & 04  33   11.46   & 14.55\\
J0854.8+2006     & S1 & 08  54  55.20 & 20   05  42.43   &  14.91\\
                 & S2 & 08  54  53.36 & 20  04   45.12   & 14.03\\
J0922.4$-$0529   & S1 & 09  22  26.71 & $-$05  29  32.06 &  15.27\\
                 & S2 & 09  22  26.88 & $-$05  29  57.14 & 14.76\\
J0927.9$-$2037   & S1 & 09  27  49.35 & $-$20  33  10.49 &  16.96\\
                 & S2 & 09  27  57.76 & $-$20  36  45.89 & 16.06\\
J1006.7$-$2159   & S1 & 10  06  45.09 & $-$21  59  22.18 &  15.45\\
                 & S2 & 10  06  33.05 & $-$21  58  29.60 & 15.25\\
J1015.0+4925     & S1 & 10  15  08.03 & 49   25  42.33   &  13.74\\
                 & S2 & 10  14  53.85 & 49  25   32.43   & 13.83\\
J1204.3$-$0708   & S1 & 12  04  06.14 & $-$07  07  56.82 &  15.41\\
                 & S2 & 12  04  32.23 & $-$07  13  24.71 & 16.12\\
J1229.1+0202     & S1 & 12  29  03.21 & 02   03  18.88   &  12.79\\
                 & S2 & 12  29  08.40 & 02  00   18.71   & 11.79\\
J1404.8+0401     & S1 & 14  04  49.21 & 04   03  37.55   &  16.22\\
                 & S2 & 14  05  00.89 & 04  00   12.71   & 16.61\\
J1512.8$-$0906   & S1 & 15  12  52.82 & $-$09  06  59.46 &  14.12\\
                 & S2 & 15  12  44.30 & $-$09  06  40.14 & 13.60\\
J1653.9+3945     & S1 & 16  53  45.82 & 39   44  09.14   &  12.49\\
                 & S2 & 16  53  28.51 & 39  46   59.20   & 13.12\\
\hline
\end{tabular}
\end{center}
\end{table*}

\begin{table*}
\caption{Log of the INOV. Columns:- (1) source name; (2) date of the observation; 
(3) Number of exposures (4) duration of the observation; (5)  and (6) {\it C}-values computed for 
the $\gamma$-ray blazars DLCs relative to the steadiest pair of comparison 
stars on any night; (7) variability status according to {\it C}-statistics, V: variable, NV: non variable; 
(8) and (9) Values of the scaled {\it F}-statistics for two blazar DLCs relative 
to two comparison stars; (10) critical $F$ value to compare with F1 and F2; (11) variability status as per {\it F}-statistics;  (12) INOV amplitude in percent; and (13) optical polarization measurements, in percent.\label{tab:stat}} 
\begin{center}
\begin{tabular}{lcccccccccccc}
\hline 
Source & Date  &  N & Duration & {\it C1}  & {\it C2} & Status  & {\it F1}  & {\it F2}  & $F_{\rm crit}$ & Status & $\psi$  & Pol.\\
       & dd mm yyyy &  &  (hr)    &              &             &             &              &              &            &  & ($\%$)  & ($\%$)\\
 (1)   & (2)      & (3)    & (4) & (5)& (6)     & (7) & (8) & (9)     &  (10)    & (11)    & (12) & (13)\\
\hline
J0608.0$-$0835	& 07 03 2015	& 15	& 1.29	& 1.87	& 2.15	& NV	& 1.57	& 2.08	& 3.69  &   NV	& 5.25	& ---	 \\
		& 03 01 2016	& 35	& 4.51	& 4.32	& 4.49	& V	& 8.47	& 9.53	& 2.26  &   V	& 8.97		& --- \\
J0656.4+4232	& 05 01 2016	& 31	& 5.06	& 1.64	& 1.92	& NV	& 0.77	& 1.06	& 2.39  &   NV	& 3.25	&  --- \\	 
J0710.5+4732	& 26 02 2015	& 59	& 2.52	& 1.89	& 2.17	& NV	& 1.36	& 2.06	& 1.86  &   NV	& 5.07	 & ---	\\ 
		& 11 03 2015	& 136	& 4.31	& 1.37	& 1.35	& NV	& 0.77	& 0.79	& 1.50  &   NV	& 2.25	& 	---\\
		& 04 01 2016	& 52	& 2.39	& 1.29	& 1.36	& NV	& 0.68	& 0.81	& 1.94  &   NV	& 3.41	& ---	\\
J0721.9+7120	& 21 01 2015	& 213	& 5.98	& 2.75	& 2.69	& V	& 6.22	& 3.98	& 1.38  &   V	& 6.17	& 9.42$\pm$0.02 \\	
		& 22 01 2015	& 92	& 4.21	& 2.86	& 2.99	& V	& 8.39	& 5.73	& 1.63  &   V	& 5.06	& 5.48$\pm$0.02 	\\
		& 23 02 2015	& 238	& 6.44	& 5.71	& 5.70	& V	& 28.90	& 22.44	& 1.35  &   V	& 11.88	& --- \\	
J0739.4+0137	& 17 03 2015	& 50	& 4.40	& 4.13	& 4.19	& V	& 8.92	& 8.54	& 1.96  &   V	& 10.07 & 4.94$\pm$0.13	\\	 
		& 02 01 2016	& 25	& 5.05	& 3.77	& 3.65	& V	& 1.38	& 1.26	& 2.72  &   NV	& 5.98	 & 0.61$^{*}\pm$0.20 (12 01 2016)	\\
J0809.8+5218	& 18 03 2015	& 28	& 4.82	& 1.34	& 1.57	& NV	& 1.10	& 1.44	& 2.46  &   NV	& 2.02	&  ---\\	 
J0831.9+0430	& 08 03 2015	& 14	& 1.27	& 4.66	& 4.55	& V	& 9.36	& 8.35	& 3.70  &   V	& 5.57	& ---	 \\
		& 09 03 2015	& 30	& 3.52	& 5.19	& 4.83	& V	& 14.60	& 11.72	& 2.39  &   V	& 13.87	& ---	\\
		& 14 03 2015	& 76	& 5.41	& 4.04	& 3.88	& V	& 7.34	& 6.52	& 1.71  &   V	& 9.17	& ---	\\
J0854.8+2006	& 10 03 2015	& 160	& 4.27	& 3.50	& 3.39	& V	& 8.53	& 12.15	& 1.45  &   V	& 9.65	& 13.38$^{*}\pm$0.07 (17 03 2015)\\
J0922.4$-$0529	& 12 03 2015	& 133	& 5.50	& 1.13	& 0.98	& NV	& 1.27	& 1.13	& 1.50  &   NV	& 4.34	& ---	\\ 
J0927.9$-$2037	& 13 03 2015	& 35	& 6.55	& 1.09	& 1.26	& NV	& 0.99	& 1.54	& 2.26  &   NV	& 3.71		& --- \\
J1006.7$-$2159	& 24 02 2015	& 48	& 8.59	& 0.95	& 1.10	& NV	& 0.72	& 1.17	& 1.99  &   NV	& 3.43	& ---	 \\
		& 15 03 2015	& 10	& 2.61	& 1.08	& 1.29	& NV	& 0.93	& 1.60	& 5.35  &   NV	& 2.21	& ---	\\
		& 16 03 2015	& 36	& 5.43	& 0.98	& 1.14	& NV	& 0.76	& 1.27	& 2.23  &   NV	& 5.24	& ---	\\
J1015.0+4925	& 20 03 2015	& 143	& 7.02	& 1.79	& 1.80	& NV	& 1.14	& 1.17	& 1.48  &   NV	& 3.86	& ---	 \\
J1204.3$-$0708	& 25 02 2015	& 36	& 6.65	& 1.17	& 1.41	& NV	& 1.34	& 1.30	& 2.23  &   NV	& 2.82	& ---	 \\
		& 17 03 2015	& 30	& 4.05	& 1.62	& 1.78	& NV	& 2.39	& 2.03	& 2.42  &   NV	& 4.24	& ---	\\
J1229.1+0202	& 26 02 2015	& 15	& 0.36	& 1.26	& 0.86	& NV	& 1.39	& 1.10	& 3.70  &   NV	& 2.30	& 0.28$^{*}\pm$0.06 (20 02 2015)	 \\
		& 10 03 2015	& 68	& 1.60	& 1.15	& 1.27	& NV	& 1.08	& 1.92	& 1.78  &   NV	& 2.53	& 0.31$^{*}\pm$0.04 (17 03 2015)	\\
		& 18 03 2015	& 300	& 4.56	& 1.20	& 1.05	& NV	& 1.07	& 1.15	& 1.31  &   NV	& 6.46	& 0.31$^{*}\pm$0.04 (17 03 2015)	\\
J1404.8+0401	& 20 05 2015	& 24	& 4.44	& 1.72	& 1.70	& NV	& 1.86	& 1.58	& 2.72  &   NV	& 3.49	& ---	 \\
J1512.8$-$0906	& 21 05 2015	& 25	& 1.89	& 1.37	& 1.24	& NV	& 1.19	& 1.28	& 2.66  &   NV	& 4.39	& 7.43$\pm$0.08	 \\
		& 22 05 2015	& 39	& 2.33	& 5.22	& 5.24	& V	& 21.48	& 31.97	& 2.16  &   V	& 15.66	& 15.32$\pm$0.05	\\
J1653.9+3945	& 20 05 2015	& 151	& 4.20	& 5.17	& 5.20	& V	& 14.11	& 12.71	& 1.46  &   V	& 11.89	& 2.20$\pm$0.05	 \\
		& 14 07 2015	& 154	& 5.41	& 1.38	& 1.36	& NV	& 0.93	& 0.83	& 1.46  &   NV	& 5.15	& 	1.28$^{*}\pm$0.04 (15 07 2015)\\
 \hline
\end{tabular}
\end{center}
\tablecomments{All polarization measurements are taken from Steward observatory (dates are quoted) and that quoted with asterisks are observed within 10 days of the INOV monitoring. See the text for details.}
\end{table*}

\begin{table*}
\caption{The results of the DCF analysis performed on the the sources whose long term \gm-ray and the optical variability behavior is shown in Figure \ref{fig_gamma1}. In the rightmost column, the name of the facilities from where the optical observations were done, are given. Note that a positive value means a lag in the optical band compared to \gm-rays.\label{tab:dcf}
}
\begin{center}
\begin{tabular}{lcl}
\hline
3FGL name  & Time (in days) & Observatory\\
\hline															     			
J0721.9+7120    & $-1.67^{+11.33}_{-10.54}$  & Steward\\
J0739.4+0137    & $+1.24^{+8.96}_{-9.30}$     & Steward\\
J0809.8+5218    & $+39.36^{+3.93}_{-56.20}$ & KAIT    \\
J0831.9+0430    & $+1.17^{+4.46}_{-4.60}$     & KAIT    \\
J0854.8+2006    & $-2.29^{+13.37}_{-3.91}$    & Steward\\
J1006.7$-$2159 & $-4.01^{+29.67}_{-5.15}$    & SMARTS\\
J1015.0+4925    & $+18.4^{+11.35}_{-13.35}$ & KAIT    \\
J1229.1+0202    & $-0.04^{+28.91}_{-4.37}$    & Steward\\
J1512.8$-$0906 & $+4.21^{+11.21}_{-10.54}$ & Steward\\
J1653.9+3945    & $-1.95^{+7.40}_{-13.95}$    & Steward\\
\hline
\end{tabular}
\end{center}
\end{table*}

\end{document}